\documentclass[aps,prd,reprint,groupedaddress,showpacs,preprintnumbers,nofootinbib,nobibnotes,amsmath,amssymb,floatfix,superscriptaddress]{revtex4-1}
\usepackage[utf8]{inputenc}
\usepackage[colorlinks=true,linkcolor=red,citecolor=blue,urlcolor=blue]{hyperref}
\usepackage{epsfig}
\usepackage{amsmath}
\usepackage{graphicx}
\usepackage{epstopdf}
\usepackage{color}
\usepackage{float}
\usepackage{xcolor}
\usepackage{amssymb}
\usepackage{enumitem}
\usepackage{caption}
\usepackage{verbatim}
\usepackage{natbib}
\usepackage[T1]{fontenc}
\usepackage{cancel}
\usepackage{natbib}
\usepackage{multirow}
\usepackage{comment}
\usepackage[utf8]{inputenc}

\definecolor{red}{rgb}{0.8,0,0}
\definecolor{violet}{rgb}{0.4,0,0.4}
\definecolor{green}{rgb}{0,0.5,0.0}
\definecolor{navy}{rgb}{0.0,0.0,0.6}
\definecolor{orange}{rgb}{0.8,0.2,0.0}
\newcommand{\vt}[1]{{\color[rgb]{0,0,0}#1}} 
\newcommand{\bp}[1]{{\textcolor{black}{#1}}}

\usepackage[normalem]{ulem}  

\begin{document}
\title{Density-induced dark-baryon conversion in $\Delta-$admixed hypernuclear neutron stars}

\author{Niyar Prabhat Kalita}
\author{Vivek Baruah Thapa}
\email{vivekthapa@bssrv.ac.in}
\affiliation{Department of Physics, Birangana Sati Sadhani Rajyik Vishwavidyalaya, Golaghat, Assam-785621}

\author{Bhanu Prakash Pant}
\affiliation{Indian Institute of Science Education and Research Tirupati, Tirupati, Andhra Pradesh–517619}

\author{Anil Kumar}
\affiliation{Institute of Theoretical Physics, Wroclaw University of Science and Technology, Wybrze\.{z}e Wyspia\'{n}skiego 27, 50-370 Wroc\l{}aw, Poland}

\author{Partha Konar}
\affiliation{Physical Research Laboratory, Ahmedabad, Gujarat-380009}

\begin{abstract} 
We investigate density-induced conversion of neutrons into a neutral dark baryon $\chi$ in cold, charge-neutral and $\beta$-equilibrated neutron-star matter containing hyperons and the complete $\Delta(1232)$ quartet. 
The hadronic sector is described within a density-dependent covariant density-functional framework using the DDME2 parameterization.
A scalar Higgs portal is also included as a possible interaction channel between the dark and hadronic sectors; its mean-field contribution is found to be numerically negligible for the couplings considered here.
\textcolor{black}{In the assumed chemical-equilibrium limit, the $\chi$ abundance is determined by the coupled equilibrium equations and total baryon-number conservation.}
We show that the appearance of hyperons and $\Delta$ resonances modifies the neutron chemical potential and delays the onset of the dark baryon, thereby reducing its abundance relative to nucleonic matter. 
This competition produces characteristic changes in the equation of state (EOS), particle composition, sound speed and adiabatic index. 
For $m_\chi=1250$, $1300$ and $1400$ MeV, the maximum masses of the complete $N+Y+\Delta+\chi$ configurations are found to be $1.80$, $1.90$ and $2.02\,M_\odot$, respectively, showing that the massive-pulsar constraint strongly disfavors the lighter
dark-baryon benchmarks for the adopted couplings. 
The radial composition further indicates that, for $m_\chi=1400$ MeV, $\chi$ is confined to the inner core of the most massive stars, while canonical configurations remain essentially insensitive to the dark sector.
The resulting stellar modifications are therefore governed predominantly by the conversion-induced rearrangement of the equilibrium composition rather than by Higgs-mediated interactions.
Our results demonstrate that the interplay between conventional non-nucleonic degrees of freedom and density-generated dark baryons is crucial for assessing the astrophysical viability of such dark-sector extensions of dense matter.
\end{abstract} 


\maketitle

\section{Introduction}
\label{sec:intro}


Compact objects, namely neutron stars (NSs) compress matter into a density regime that cannot be reproduced in terrestrial laboratories \cite{1996cost.book.....G}. Their cores may reach several times the nuclear saturation density ($n_0$), where the composition of matter is controlled by a balance among strong interactions, charge neutrality, weak equilibrium and the rapidly increasing chemical potentials of the constituent particles.
Although the global behavior of a NS is governed by the EOS of this matter, the relevant density domain is not yet directly accessible through first-principles calculations of quantum chromodynamics. The microscopic composition of the stellar core therefore remains uncertain \cite{Lattimer2016PhR, 2007PrPNP..59...94W,Sedrakian2007PrPNP}.

The range of acceptable EOSs has nevertheless narrowed considerably through observations.
The existence of pulsars with masses close to $\sim2~M_\odot$ \cite{2010Natur.467.1081D,Arzoumanian_2018, Antoniadis_2013, 2020NatAs...4...72C} requires sufficient pressure at high density, while the mass-radius measurements reported by NICER \cite{2019ApJ...887L..24M, 2019ApJ...887L..21R} and the tidal response  inferred from binary NS mergers \cite{LIGO_Virgo2017b, LIGO_Virgo2017c, LIGO_Virgo2017a, PhysRevX.9.011001} constrain the behavior of matter at intermediate densities.
These different measurements do not by themselves identify the particles present in the core, but they place strong restrictions on any new degree of freedom that appreciably lowers the pressure. A viable microscopic model must therefore accommodate the emergence of additional species without losing consistency with the observed masses, radii and tidal deformabilities of NSs.

Among the possible non-nucleonic constituents of NS matter, strange baryons and spin-(3/2) $\Delta-$resonances have received sustained attention. Hyperons become energetically favourable when the baryonic chemical potentials exceed their in-medium threshold energies, allowing part of the nucleonic Fermi pressure to be released through the population of additional baryonic states \cite{1985ApJ...293..470G, 1995PhRvC..52.3470K, 1997NuPhA.625..435B}.
Their appearance, however, generally lowers the pressure at a given energy density and may drive the maximum stellar mass below the observationally established $\sim2~M_\odot$ scale.
The resulting tension between microscopic expectations and pulsar-mass measurements constitutes the well-known hyperon puzzle \cite{Weissenborn2012a, Chatterjee2015, Oertel2015, Oertel_RMP_2017}.
Within covariant density-functional (CDF) approaches, this difficulty may be alleviated by sufficiently repulsive hyperon-vector interactions, additional hidden-strangeness channels, or an appropriate density dependence of the meson-baryon couplings \cite{Bonanno2012A&A, Colucci_PRC_2013, 2016PhRvC..94c5804F}.
The role of the $\Delta$ quartet is comparably subtle because its threshold densities depend strongly on the poorly constrained scalar and vector $\Delta$-meson couplings.
Studies \cite{Drago_PRC_2014, Cai_PRC_2015, Zhu_PRC_2016} show that the negatively charged $\Delta^{-}$ may appear at only a few times $n_0$.
Calculations incorporating both hyperons and $\Delta$ resonances further indicate that the $\Delta$ sector can appreciably decrease the radii of intermediate-mass stars while producing a more modest change in the maximum mass \cite{Thapa_2021, Parmar_2025}.

These compact stars may also probe weakly coupled particles belonging to a dark sector.
Many past works \cite{1989PhRvD..40.3221G, PhysRevD.77.023006, 2008PhRvD..77d3515B, 2021PhRvD.103d3019G} have assumed that dark matter (DM) is gravitationally captured from the Galactic environment and subsequently thermalized within the star.
Depending on whether the dark particles form a compact core, an extended halo, or a component mixed with the baryonic fluid, their presence can affect the mass-radius relation, tidal deformability, cooling evolution and oscillation spectrum
\cite{2011PhRvD..84j7301L, 2019PhRvD..99d3016D, 2021MNRAS.507.4053D, 2024JCAP...12..042T, 2025MNRAS.544.3549S}.

In several phenomenological descriptions of DM-admixed NSs as in Refs.\cite{2017PhRvD..96h3004P, 2019PhRvD..99d3016D, 2020MNRAS.495.4893D, 2018PhRvD..97l3007E, 2021MNRAS.507.4053D}, the dark component is introduced by prescribing a fixed Fermi momentum or total mass fraction, rather than deriving its abundance from chemical equilibrium with the baryonic medium.
The resulting dark population is therefore a model input and is not fixed by the local baryonic chemical potentials or by an in-medium production threshold. The capture of DM through scattering by NSs, under standard Galactic-halo conditions, is generally insufficient to generate substantial changes in the NS observables
\cite{2019JCAP...06..054B, 2020JCAP...09..028B, 2024PhR..1052....1B}. 
\textcolor{black}{Recent single-fluid developments also allow the local dark density to vary with the baryonic density, rather than prescribing a constant dark Fermi momentum \cite{Issifu_2026, issifu2026selfconsistenthiggsportalframeworkdark}. These constructions should be distinguished from the present conversion-equilibrium prescription, in which neutron and $\chi$ numbers are not separately conserved and their chemical potentials satisfy \(\mu_n=\mu_\chi\).}

Several studies \cite{kumar2026slowlyrotatingtwofluidneutron, issifu2026darkmatterheatingevolving, das2020darkmatteradmixedneutron,Mariani_2023} suggest a two-fluid approach of DM-admixed NSs, where the dark components are assumed to form a fluid co-existing with the baryonic sector. Dark fluid interacts with the baryons only through gravity \cite{issifu2026darkmatterheatingevolving,das2020darkmatteradmixedneutron,kumar2026slowlyrotatingtwofluidneutron}. The central energy density as well as pressure of both the sectors evolve separately, and the stellar structure and the other properties of the NS follow as a combined consequence. This approach is useful in describing the effect of DM when it is accumulated in NS from the environment as discussed already, but does not address the question of the origin of the DM.

A distinct route to a dark component in NSs is its direct production from baryonic matter.
This possibility was highlighted by Fornal and Grinstein \cite{Fornal_2018}, who proposed nonstandard neutron-decay channels such as $n\rightarrow\chi+\gamma$ and $n\rightarrow\chi+\Phi$ in connection with the beam–bottle neutron-lifetime discrepancy.
Here, $\chi$ is a neutral spin-(1/2) state carrying unit baryon number, while $\Phi$ denotes a light dark-sector boson.
Refs.\cite{2018PhRvL.121f1801B, 2018PhRvL.121f1802M} point that in dense matter, the neutron chemical potential can rise well above its vacuum mass, allowing an in-medium conversion into a dark baryon even when the corresponding vacuum process is suppressed.
Over the long lifetime of the star, provided that the conversion reaches equilibrium, this mechanism may still produce an appreciable $\chi$ population.
More recently, neutron--dark-sector conversion in dense, $\beta$-equilibrated matter has been investigated from the perspectives of the NS EOS, nuclear symmetry energy, and astrophysical constraints \cite{2024PhRvD.110h3003B, 2026PhRvC.113e5807D}.
We stress that the mass range considered here,
$m_\chi=1.25$--$1.40$ GeV, lies above the neutron mass. Thus, $\chi$ is not intended to realize the free-neutron lifetime-anomaly scenario, but is instead treated as a heavier dark-sector baryon that
can become populated only in sufficiently dense matter.
The cosmological and astrophysical viability of baryon-number carrying dark states, including their stability and possible
decay channels, is generally model dependent
\cite{2021PhRvD.103k5002M}.

In the present work, we investigate density-induced dark-baryon conversion in cold, charge-neutral, and $\beta-$equilibrated $\Delta$-resonance admixed hypernuclear matter within the framework of CDF theory. The EOS is constructed using the DDME2 parameterization with density-dependent meson-baryon couplings \cite{2005PhRvC..71b4312L}. 
This framework has previously been extended to hypernuclear matter containing the complete $\Delta$ quartet \cite{Thapa_2021, Parmar_2025}, where the resulting equations of state were found to remain broadly consistent with the available astrophysical constraints, and has also been employed to study the influence of strong magnetic fields \cite{particles3040043}. 
In the present extension, a neutral dark baryon $\chi$ is generated through in-medium neutron conversion \cite{Fornal_2018, 2018PhRvL.121f1802M}. 
A scalar Higgs portal is retained as a benchmark interaction channel between $\chi$ and the hadronic sector \cite{2019PhRvD..99d3016D}.
The novel aspect of the present study is the self-consistent competition among the density-generated dark baryon, baryon-octet and $\Delta$ quartet. In particular, we determine how the prior appearance of conventional non-nucleonic species modifies the threshold and abundance of $\chi$, and how the resulting rearrangement of the stellar composition affects the EOS and observable NS properties.
\textcolor{black}{Related studies have explored the effects of a bosonic sexaquark dark-matter candidate on the structure and oscillation properties of neutron stars containing hyperonic and quark matter \cite{Shahrbaf_2026_probing,Shahrbaf_2026_observational}. 
These investigations provide a complementary perspective on the interplay between dark-sector particles and conventional dense-matter degrees of freedom.
The present study instead considers a fermionic dark baryon populated through neutron conversion in $\Delta$-admixed hypernuclear matter.}


The remainder of the paper is organized as follows. 
Sec. \ref{sec:formalism} presents the density-dependent hadronic model and the neutron–dark-baryon conversion framework. The Higgs-portal extension and a quantitative estimate of its contribution are provided in Appendix A.
The numerical results and their respective discussions are provided in Sec. \ref{sec:results}. Sec. \ref{sec:conclusions} summarizes the principal conclusions and outlines possible extensions of the framework. 
Natural units, $\hbar=c=1$, are used throughout.

\section{Formalism} \label{sec:formalism}
\subsection{Density Dependent Hadronic CDF Model}
\label{CDF}
This section summarizes a density-dependent model for investigating the appearance of \(\Delta-\)resonances, \(\Lambda\), \(\Xi\) hyperons. 

\bp{The NS matter is assumed to be composed of baryon octet (\(N\), \(\Lambda\), \(\Xi\), \(\Sigma\)), \(\Delta-\)resonances (\(\Delta^-\), \(\Delta^0\), \(\Delta^+\), \(\Delta^{++}\)), and the proposed dark baryon \(\chi\).}
\vt{We distinguish between the interacting hadronic density,
$n_{\rm H}=\sum_{j= B,\Delta} n_j$,
and the total conserved baryon density,
$n_b=n_{\rm H}+n_\chi$.
Since $\chi$ does not couple directly to the density-dependent
$\sigma$, $\omega$, $\rho$ and $\phi$ vertices, the coupling functions of the DDME2 functional are evaluated at $n_{\rm H}$. The dark baryon nevertheless contributes to the total baryon number, energy density, pressure and gravitational mass.}
The already established RMF \cite{1996cost.book.....G} provides a mechanism for strong interactions among nucleons, and an extension of the framework corresponding to \(\Delta-\)resonances, \(\Lambda\) and \(\Xi\) is explored in Refs. \cite{Thapa_2021, Parmar_2025}.

A scalar meson field \(\sigma\) mediates the attractive strong interaction, whereas the vector meson \(\phi_\mu\) accounts for the interactions involving strange particles. The isoscalar-vector \(\omega_\mu\) and the isovector-vector \(\rho_\mu\) govern the repulsive nature and isospin structure, respectively.
The meson–hadron coupling vertices depend on \(n_H\); the mean fields are determined self-consistently from the field equations.
The hadronic Lagrangian density is given by \cite{Thapa_2021},
\begin{align}
    \mathcal{L}_{\rm had} &= \mathcal{L}_B+\mathcal{L}_{\Delta}+\mathcal{L}_l+\frac{1}{2}\left(\partial_\mu \sigma\partial^\mu \sigma-m^2_{\sigma}\sigma^2\right) \notag\\
    &\quad -\frac{1}{4}\omega_{\mu\nu}\omega^{\mu\nu}+\frac{1}{2}m_{\omega}^2\omega_\mu\omega^\mu-\frac{1}{4}\rho_{\mu\nu}\rho^{\mu\nu} \notag\\
    &\quad +\frac{1}{2}m_{\rho}^2\rho_\mu\rho^\mu-\frac{1}{4}\phi_{\mu\nu}\phi^{\mu\nu}+\frac{1}{2}m_{\phi}^2\phi_\mu\phi^\mu,
    \label{hadron}
\end{align}
\textcolor{black}{where \(\mathcal L_B\) and \(\mathcal L_l\) describe the spin-\(1/2\) baryon octet and leptons, respectively, while \(\mathcal L_\Delta\) describes the spin-\(3/2\) $\Delta-$quartet using Rarita–Schwinger vector-spinor fields}, given by,
\begin{subequations}
\begin{eqnarray}
    \mathcal{L}_B&=&\sum_{B} \bar{\psi}_B(i\gamma^\mu D_{\mu {(B)}} - m_B^*)\psi_B, \label{dirac} \\
    \mathcal{L}_l&=&\sum_l \bar{\psi}_l(i\gamma^\mu \partial_\mu - m_l)\psi_l, \label{dirac_lep} \\
    \mathcal{L}_\Delta&=&\sum_{\Delta} \bar{\psi}^\nu_{\Delta}(i\gamma^\mu D_{\mu_{(\Delta)}} - m_\Delta^*)\psi_{\Delta \nu}, \label{delta}    
\end{eqnarray}
\end{subequations}
where, \(D_{\mu (j)}=\partial_\mu+ig_{\omega j}\omega_{\mu}+ig_{\rho j}\tau_j.\rho_\mu +i g_{\phi j}\phi_\mu\) is a covariant derivative, where \(j\) represents (\(B,\Delta\)).
\textcolor{black}{Eq.\eqref{delta} is used in the uniform-matter approximation restricted to the physical spin-(3/2) sector, following Ref. \cite{Thapa_2021}. Only the four physical spin states of each $\Delta$ charge state are included in the quasiparticle sums; the additional components of the vector-spinor field are not counted as independent particle states.
The Dirac fields of the baryon octet and leptons are denoted by $\psi_B$, $\psi_l$ respectively, while the Rarita-Schwinger fields of the $\Delta-$quartet are denoted by $\psi_{\Delta \nu}$.}
It is to be noted that $g_{\phi j}=0$ for nucleons and $\Delta$ resonances.
\(\tau_j\) is the isospin operator for \(\rho_\mu\) meson corresponding to \(j\). \(g_{ij}\) are density-dependent coupling constants, where \(i\) represents meson fields. The hadronic effective masses in Eq.\eqref{dirac} are given by
\begin{align}
    m_j^*=m_j-g_{\sigma j}\sigma.
    \label{effective_mass}
\end{align}
The scalar and vector densities of baryons are given by \(n^s=\langle \bar\psi\psi \rangle\) and \(n_{\rm H}=\langle \bar\psi\gamma^0\psi \rangle\) respectively. In density-dependent models, coupling parameters are functions of density \cite{Parmar_2025,Malik_2022,Malikk_2022,Beznogov_2023}, as mentioned below
\textcolor{black}{
\begin{align}
    g_{iN}(n_{\rm H})=g_{iN}(n_0)f_i(x)
\end{align}
where, \(x=n_{\rm H}/n_0\)} and 
\begin{align}
    f_i(x)=a_i\frac{1+b_i(x+d_i)^2}{1+c_i(x+d_i)^2}
\end{align}
where \(i=\sigma, \omega\) mesons, and for \(\rho\) mesons, it is given by
\begin{align}
f_i(x)=\exp(-a_\rho(x-1))   .
\end{align}

\textcolor{black}{All density-dependent hadronic vertices, including those obtained through fixed hyperon-to-nucleon and $\Delta$-to-nucleon coupling ratios, are evaluated at \(n_H\). At each prescribed total density \(n_b\), the hadronic density \(n_H=n_b-n_\chi\) is determined self-consistently. The derivatives entering the rearrangement self-energy (Eq.\eqref{rearrangement}) are taken with respect to this same density.}
The values of the parameters involved in density-dependent couplings are taken according to the parameterization of DDME2 (see Table-\ref{tab:ddme2}). The mass of the nucleons is taken to be $938.9$ MeV, and that of $\Lambda$, $\Xi^-$, $\Xi^0$, and the $\Delta$-resonances are considered $1115.68$ MeV, $1321.71$ MeV, $1314.86$ MeV and $1232.0$ MeV, respectively.
\begin{table*}[t]
\centering
\caption{DDME2 density-dependent coupling parameters ($m_\sigma$,$m_\omega$ and $m_\rho$ are in MeV)}
\label{tab:ddme2}
\resizebox{0.9\textwidth}{!}{%
\begin{tabular}{lccccccccccccc}
\hline
\hline
$m_\sigma$ (MeV) & $m_\omega$ (MeV) & $m_\rho$ (MeV) &
$a_\sigma$ & $a_\omega$ & $a_\rho$ & $b_\sigma$ & $b_\omega$ & $c_\sigma$ & $c_\omega$ & $d_\sigma$ & $d_\omega$ \\
\hline
550.1238 & 783 & 763 &
1.3881 & 1.3892 & 0.5647 & 1.0943 & 0.9240 & 1.7057 & 1.4620 & 0.4421 & 0.4775 \\
\hline
\hline
\end{tabular}}
\end{table*}

The expressions for field densities, pressure, energy density, and other relevant quantities are well-discussed in \cite{Thapa_2021,Parmar_2025,Malik_2022}.
In the ground state of uniform nuclear matter, only the temporal components of the vector fields survive. The corresponding mean-field equations are
\begin{align}
m_\sigma^2\sigma
&=
\sum_{j= B,\Delta} g_{\sigma j}n_j^s,\\
m_\omega^2\omega_0
&=
\sum_{j= B,\Delta} g_{\omega j}n_j,\\
m_\rho^2\rho_{03}
&=
\sum_{j= B,\Delta} g_{\rho j}I_{3j}n_j,\\
m_\phi^2\phi_0
&=
\sum_{j= B,\Delta} g_{\phi j}n_j ,
\end{align}
where $n_j$ and $n_j^s$ denote the vector and scalar densities,
respectively. The scalar density is
\begin{equation}
n_j^s
=
\frac{\gamma_j}{2\pi^2}
\int_0^{k_{Fj}}
\frac{m_j^* k^2\,dk}{\sqrt{k^2+m_j^{*2}}},
\end{equation}
\textcolor{black}{where the spin degeneracy is \(\gamma_j=2J_j+1\), with \(\gamma_j=2\) for each spin-\(1/2\) species and \(\gamma_j=4\) for each $\Delta$ charge state. The charge states are summed separately, so no additional isospin multiplicity is included. The corresponding number density is \(n_j=\gamma_j k_{Fj}^{3}/(6\pi^2)\)}, and $k_{F_j}$ is the Fermi momentum of the $j-$th baryon.
\textcolor{black}{The numerical implementation consistently uses $\gamma_\Delta=4$ for each separately treated $\Delta$ charge state in the number density, scalar density, energy density and pressure.}
The density dependence of the meson-baryon vertices gives rise to the rearrangement self-energy
\begin{equation}
\begin{aligned}
\Sigma^R
=&
\sum_{j= B,\Delta}
\bigg[
\frac{\partial g_{\omega j}}{\partial n_{\rm H}}\omega_0 n_j
+
\frac{\partial g_{\rho j}}{\partial n_{\rm H}}
I_{3j}\rho_{03}n_j
+
\frac{\partial g_{\phi j}}{\partial n_{\rm H}}\phi_0n_j \\ &
-
\frac{\partial g_{\sigma j}}{\partial n_{\rm H}}\sigma n_j^s
\bigg].
\end{aligned}
\label{rearrangement}
\end{equation}
This contribution is required to preserve thermodynamic consistency in the density-dependent covariant functional.
\vt{The chemical potential of a hadronic baryon $j$ is,
\begin{equation}
\mu_j = \sqrt{k_{Fj}^2+m_j^{*2}} + g_{\omega j}\omega_0
+ g_{\rho j}I_{3j}\rho_{03}
+ g_{\phi j}\phi_0
+ \Sigma^R .
\end{equation}
For the leptons,
$\mu_l=\sqrt{k_{Fl}^2+m_l^2},
\qquad l=e,\mu$.}

\subsection{Coupling parameters}


A detailed discussion on meson-hyperon coupling parameters is performed in \cite{Thapa_2021}. Using SU(6) symmetry and the quark counting rule, the following relations are obtained:
\begin{align}
g_{\omega\Lambda} &= g_{\omega\Sigma} = \frac{2}{3}\, g_{\omega N}, 
\qquad g_{\omega\Xi} = \frac{1}{3}\, g_{\omega N}, \\[4pt]
g_{\phi\Lambda} &= g_{\phi\Sigma} = -\frac{\sqrt{2}}{3}\, g_{\omega N}, 
\qquad g_{\phi\Xi} = -\frac{2\sqrt{2}}{3}\, g_{\omega N}, \\[4pt]
g_{\rho\Sigma} &= 2\, g_{\rho\Xi} = g_{\rho N}, 
\qquad g_{\rho\Lambda} = 0.
\end{align}

The meson-$\Delta$ baryon coupling values are not known due to little experimental exploration. We treat them as free parameters as done in Ref.-\cite{Thapa_2021}. The values for scalar meson-hyperon couplings are fixed by taking the hyperon optical potentials, with values as \(U_\Lambda = -30\) MeV, \(U_\Sigma = +30\) MeV, and \(U_\Xi = -14\) MeV \cite{Thapa_2021,Parmar_2025}. We take the values of meson-nucleons coupling values according to DDME2 parameterization, as done in the mentioned works-

\begin{equation*}
    g_{\sigma N} (n_0)=10.5396,
    \qquad 
    g_{\omega N} (n_0)=13.0189,
\end{equation*}
\vspace*{-0.7 cm}
\begin{equation}
    g_{\rho N} (n_0)=7.3672
\end{equation}
\hspace*{2 mm}The meson-$\Delta$ coupling ratios are defined as
\begin{equation}
R_{\sigma\Delta} = \frac{g_{\sigma\Delta}}{g_{\sigma N}}, \qquad
R_{\omega\Delta} = \frac{g_{\omega\Delta}}{g_{\omega N}}, \qquad
R_{\rho\Delta} = \frac{g_{\rho\Delta}}{g_{\rho N}}.
\end{equation}


\textcolor{black}{The $\Delta$–meson couplings remain uncertain. Ref.\cite{Lopes_2023} develops a symmetry-based coupling scheme using SU(3)/SU(6) relations together with baryon-potential constraints. Here we retain phenomenological coupling ratios to explore this uncertainty. We fix \(R_{\omega\Delta}=1.1\) and \(R_{\rho\Delta}=1.0\), adopt \(R_{\sigma\Delta}=1.2\) for the benchmark calculations, and vary \(R_{\sigma\Delta}\) in the parameter study.}


\subsection{Dark-Baryon Conversion Model}
\label{DM-Model}

Our model introduces an in-medium conversion channel between neutrons and a neutral baryon-number-carrying dark-sector fermion $\chi$ of spin $\frac{1}{2}$, motivated by Refs.~\cite{Fornal_2018, 2018PhRvL.121f1802M}. 
The parameter \(g_c\) controls the strength of the off-diagonal neutron-dark-baryon mixing, 
\begin{align}
    \mathcal{L}_{\rm conv}=g_c(\bar{n}\chi+\bar{\chi}n).
    \label{conversion}
\end{align}

\textcolor{black}{The parameter \(g_c\) is an off-diagonal neutron–dark-baryon mixing amplitude with dimensions of energy. 
Here \(n\) and \(\chi\) denote the neutron and dark-baryon Dirac fields, respectively, with adjoints \(\bar n=n^\dagger\gamma^0\) and \(\bar\chi=\chi^\dagger\gamma^0\). In natural units, both fields have mass dimension \(3/2\), while the real mixing amplitude \(g_c\) has mass dimension \(1\), ensuring that \(\mathcal L_{\mathrm{conv}}\) has mass dimension \(4\).
The interaction in Eqn. \eqref{conversion} has the same structure as those considered in Refs.~\cite{Fornal_2018, 2018PhRvL.121f1802M}. In particular, McKeen et al. \cite{2018PhRvL.121f1802M} discuss in-medium conversion for dark-baryon masses above the vacuum neutron mass and provide an approximate argument that very small mixing can permit equilibration over stellar timescales. Their estimate accounts for neutron self-energy effects on the effective mixing, although a detailed production-rate calculation is deferred. Complementarily, Berezhiani et al. \cite{2021EPJC...81.1036B} investigate collision-assisted neutron–mirror-neutron conversion and illustrate how in-medium mixing and Pauli blocking govern the production rate. Their analysis also shows that conversion slows as the mirror population grows, emphasizing the importance of accounting for the evolving composition when assessing equilibration.
These studies provide physical motivation for considering the weak-mixing chemical-equilibrium limit. However, they do not quantitatively establish its validity for our heavier dark-baryon benchmarks and density-dependent hypernuclear matter. We impose \(\mu_n=\mu_\chi\) and neglect mixing-induced corrections to the quasiparticle spectra and thermodynamic potential. The simultaneous validity of these approximations requires an equilibration time shorter than the relevant stellar evolution timescale and spectral and thermodynamic corrections below the accuracy required for the equation of state. Establishing an allowed interval of \(g_c\) requires an in-medium rate calculation, including occupation factors and reverse reactions, together with an assessment of the mixing-induced corrections. We have not established this interval; the stellar configurations presented here are therefore conditional on the assumed weak-mixing chemical-equilibrium limit.}
Within this assumed equilibrium limit, the onset of a nonzero dark-baryon population is determined by the energy required to populate the lowest available ($\chi$) state. Analogously to the threshold construction for additional baryonic species \cite{1985ApJ...293..470G, 1995PhRvC..52.3470K, 1997NuPhA.625..435B}, the thermodynamic onset condition is,

\begin{align}
    \mu_n\geqslant m_\chi^*,
    \label{onset1}
\end{align}
where $\mu_n$ is the neutron chemical potential and $m_\chi^*$ is the in-medium dark-baryon mass. In the Higgs-extended calculation, $m_\chi^*=m_\chi-y_\chi h$ where, $y_\chi$ is the Yukawa coupling term which determines the strength of coupling of DM with the Higgs field $h$. We find numerically that the corresponding Higgs-induced correction is negligible for the adopted couplings, such that $m_\chi^*\simeq m_\chi$ to extremely high accuracy. The Higgs-sector formalism and the numerical estimate of this correction are given in Appendix A.
Eqn.~(\ref{onset1}) determines only the threshold for the first appearance of $\chi$. At the onset, $k_{F\chi}=0$ and $\mu_n=m_\chi^*$. 
\textcolor{black}{This condition identifies when populating \(\chi\) becomes thermodynamically favourable; it does not specify an elementary conversion process or its rate. The latter additionally requires a reaction mechanism satisfying energy–momentum conservation and the occupation constraints of the medium.}
Above the threshold, the equilibrium abundance is
obtained by imposing
\begin{equation}
\mu_n=\mu_\chi ,
\end{equation}
where
$\mu_\chi=\sqrt{k_{F\chi}^2+m_\chi^{*2}}$.
The dark-baryon number density is given by,
\begin{eqnarray}
\begin{aligned}
n_\chi&=\frac{k_{F\chi}^3}{3\pi^2}, \label{22a} \\
n_\chi&=
\begin{cases}
\dfrac{(\mu_n^2-m_\chi^{*2})^{3/2}}{3\pi^2},
&\mu_n>m_\chi^*,\\[6pt]
0,&\mu_n\le m_\chi^*.
\end{cases}
\end{aligned}
\end{eqnarray}
\textcolor{black}{At fixed neutron chemical potential, a smaller dark-baryon mass produces a larger equilibrium $\chi$ density. At fixed total baryon density, however, the neutron chemical potential also changes through the coupled composition and mean-field equations. The ordering of the EOSs therefore follows from the complete self-consistent solution.}

The complete system is subject to total baryon-number conservation, charge neutrality and weak equilibrium \cite{Thapa_2021}:
\begin{align}
n_b
&=
n_\chi+\sum_{j= B,\Delta}n_j,\\
\sum_{j= B,\Delta}&q_jn_j-n_e-n_\mu
=0,\\
\mu_j
&=
\mu_n-q_j\mu_e,\\
\mu_\mu&=\mu_e
\end{align}
where, \(\mu_j\) and \(q_j\) are the chemical potential and charge of the \(j^{th}\) baryon respectively, and \(\mu_e\) is the chemical potential of electron.
These relations are solved simultaneously with the meson and Higgs mean-field equations at every value of $n_b$.
In addition to the conversion mechanism, we have examined a scalar Higgs portal as a possible interaction channel between the dark and hadronic sectors, and the
corresponding formalism and its quantitative assessment are explained in Appendix A and only the leading equilibrium dark-baryon dynamics is discussed below.


In the present phenomenological framework, $\chi$ is assumed to be an
electrically neutral spin-$1/2$ fermion carrying baryon number
$B_\chi=1$. These quantum numbers allow neutron--$\chi$ conversion
without violating electric-charge or baryon-number conservation.
For a light dark baryon satisfying
$m_\chi<m_p+m_e\simeq938.78$ MeV, the decay
$\chi\rightarrow p+e^-+\bar{\nu}_e$ is kinematically forbidden, so
that $\chi$ can be stable in the absence of additional decay channels
and may consequently constitute a dark-matter candidate
\cite{Fornal_2018}. The situation is different for the mass
range considered here, $m_\chi=1.25$--$1.40$ GeV, which lies above
the neutron mass. In this regime, baryon-number and electric-charge
conservation alone do not guarantee the vacuum stability of $\chi$;
its lifetime depends on the underlying dark-sector interactions and
the allowed decay channels. If cosmological stability is required,
an additional protecting symmetry or sufficiently suppressed
interactions must therefore be invoked
\cite{2021PhRvD.103k5002M}. We consequently interpret $\chi$ here
as a heavier baryon-number-carrying dark-sector fermion produced in
dense NS matter, without assuming that it constitutes the dominant cosmological dark matter.
It is also noteworthy that NS mass constraints have been shown to strongly disfavor weakly interacting dark baryons near the nucleon mass, while masses above approximately $1.2$ GeV provide a
possible route to avoiding excessive softening in the absence of strong dark-sector repulsion \cite{2018PhRvL.121f1802M}.
The free dark-baryon sector is described by \cite{2019PhRvD..99d3016D},
\begin{align}
    \mathcal{L}_{\chi}=\bar{\chi}[i\gamma^\mu\partial_\mu-m_\chi]\chi
    \label{chi}
\end{align}
where, \(m_\chi\) is the mass of \(\chi\).\\
\hspace*{4 mm} The full Lagrangian density is as follows,
\begin{align}
    \mathcal{L}=\mathcal{L}_{\rm had}+\mathcal{L}_{\chi}+\mathcal{L}_{\rm conv}+\mathcal{L}_{\rm Higgs},
\end{align}
where ${\cal L}_{\rm had}$, ${\cal L}_{\chi}$ and ${\cal L}_{\rm conv}$ are defined above, while the Higgs-portal term ${\cal L}_{\rm Higgs}$ and its numerical contribution are discussed in Appendix A.
\textcolor{black}{The energy density and pressure below are evaluated using unmixed quasiparticle spectra, with chemical equilibrium imposed as a constraint on the composition. For nondegenerate unmixed states, the off-diagonal interaction gives no first-order energy correction, so the leading spectral shifts are quadratic in \(g_c\) and are assumed negligible here. Consequently, \(g_c\) does not enter the EOS explicitly at the retained order, although it controls the conversion kinetics. Negligible mixing corrections and equilibration within stellar evolution timescales remain separate requirements, whose simultaneous validity is assumed rather than quantitatively demonstrated in this work.}

\textcolor{black}{At leading order in the neglected mixing and Higgs corrections, the energy density is,}

\textcolor{black}{
\begin{align}
\varepsilon &= \frac{1}{2}m_\omega^2\omega_0^2 + \frac{1}{2}m_\sigma^2\sigma^2 + \frac{1}{2}m_\rho^2\rho_3^2 + \frac{1}{2}m_\phi^2\phi_0^2 \notag\\
&\quad + \sum_{j=B,l,\chi} \frac{\gamma_j}{2 \pi^2} \int_{0}^{k_{F_j}} k^2\sqrt{k^2+(m_{j}^*)^2} \,dk \notag\\
&\quad + \sum_{\Delta} \frac{\gamma_\Delta}{2 \pi^2} \int_{0}^{k_{F_\Delta}} k^2\sqrt{k^2+(m_{\Delta}^*)^2} \,dk.
\label{epstotal}
\end{align}}
For \(e^-\) and \(\mu^-\), $m_j^*=m_j$ (bare masses), while for 
other particle species,
\vt{the in-medium effective masses are
\begin{align}
m_j^*
&=
m_j-g_{\sigma j}\sigma,
\qquad j= B,\Delta,\\
m_\chi^*&\simeq m_\chi .
\label{effective_chi}
\end{align}}
\textcolor{black}{At zero temperature, the pressure follows from the thermodynamic Euler relation,
\begin{align}
P=\sum_{j=B,\Delta}\mu_jn_j+\sum_{l=e^-,\mu^-}\mu_ln_l+\mu_\chi n_\chi-\varepsilon,
\label{Pressure}
\end{align}
where the sum includes every populated hadronic species, the leptons and the dark baryon. The hadronic chemical potentials include the rearrangement contribution in Eq. \eqref{rearrangement}.
}
The effect of the scalar portal was assessed by solving the Higgs-extended mean-field equations.
For the benchmark coupling $y_{\chi}=0.01$, the Higgs-induced effective-mass correction is of order $10^{-9}$ MeV and therefore has no discernible influence on the EOS or stellar observables; a quantitative estimate is provided in Appendix A.
 


At each prescribed total baryon density, the coupled hadronic mean-field, charge-neutrality, baryon-number conservation and chemical-equilibrium equations are solved simultaneously. The full calculation additionally includes the Higgs mean-field equation as
described in Appendix A.
The core EOS is matched to the Baym, Pethick $\&$ Sutherland \cite{1971ApJ...170..299B} crust EOS, with continuity imposed in pressure and chemical potential following Fortin et al. \cite{2016PhRvC..94c5804F}.
The stellar calculation uses an effective one-fluid barotropic description of the locally equilibrated mixture. Relative motion between the hadronic and dark components is not treated. The imposed chemical-equilibrium condition does not, by itself, establish rapid momentum exchange between the components.
The resulting barotropic EOS is used to integrate
the Tolman--Oppenheimer--Volkoff (TOV) equations \cite{1996cost.book.....G}. 
The tidal Love number is
obtained by integrating the corresponding first-order metric
perturbation equation together with the stellar structure equations.
All EOS tables are checked for monotonicity, thermodynamic stability,
and causality.

\section{Results} \label{sec:results}

The principal objective of this section is to examine how the
density-induced appearance of the dark baryon $\chi$ modifies the
microscopic composition and macroscopic properties of NSs.
Particular emphasis is placed on the competition among $\chi$,
hyperons, and the $\Delta(1232)$ resonances. 
For the benchmark calculations, we consider three representative dark-baryon masses, $m_\chi=1250$, $1300$, and $1400$ MeV, while fixing $R_{\sigma\Delta}=1.2$, $R_{\omega\Delta}=1.1$ and $R_{\rho\Delta}=1.0$. The Higgs-portal benchmark $y_\chi=0.01$ is retained in the complete numerical calculation (see Appendix A).
The combined dependence on $m_\chi$ and $R_{\sigma\Delta}$ is
subsequently investigated through a two-dimensional parameter scan.

\subsection{Dark-baryon threshold and chemical equilibrium}
\label{subsec:chemical_potential}

\begin{figure}[H]
    \centering
    \includegraphics[width=1.0\linewidth]{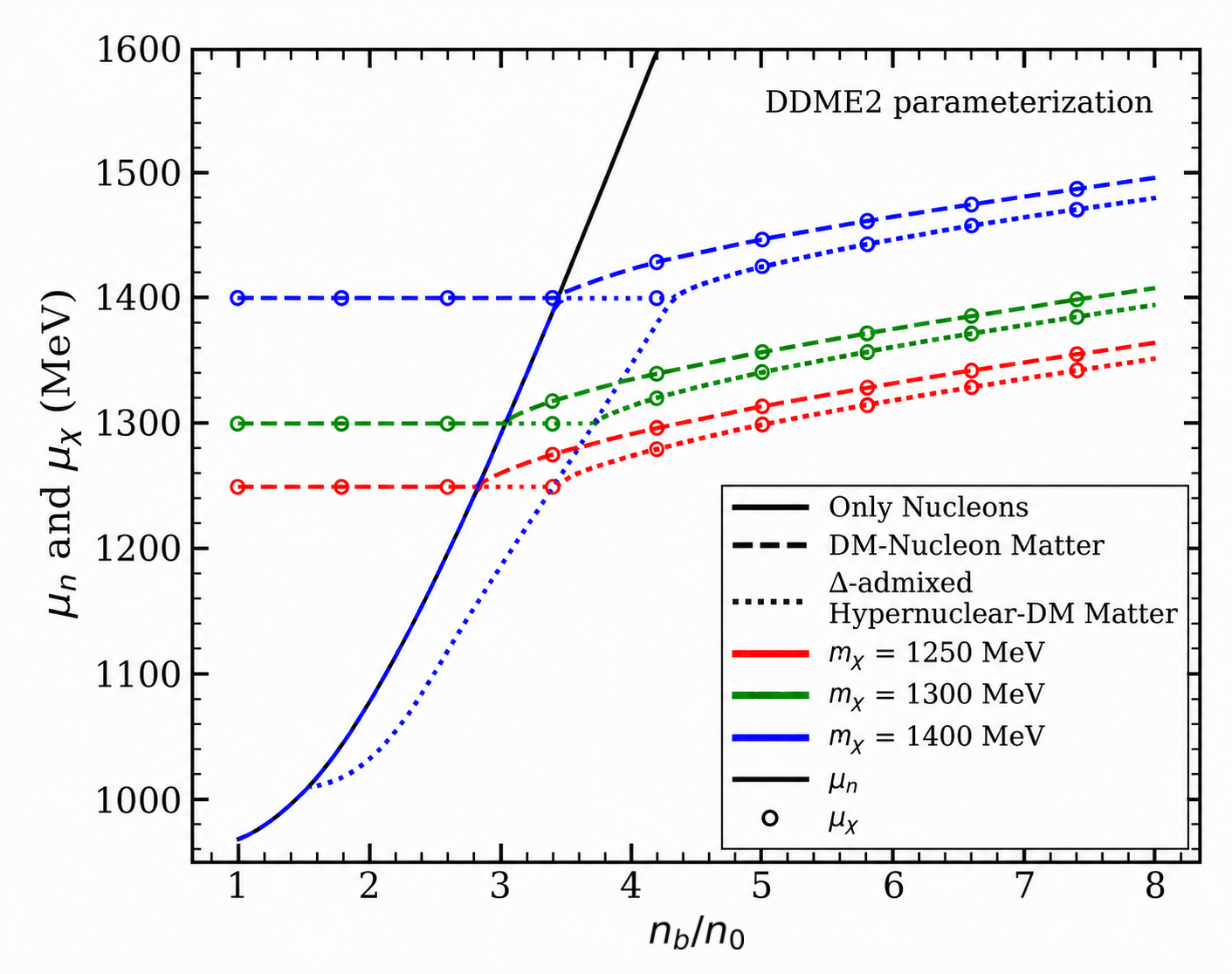}
    \caption{Chemical potentials of neutron $(\mu_n)$ and DM $(\mu_\chi)$ as functions of number density $n_b/n_0$ for NS matter composition: (i) Only Nucleons, (ii) DM-Nucleon matter and, (iii) $\Delta$-admixed Hypernuclear Matter for the masses of DM: $M_{\chi}=1250, 1300$ and $1400$ MeV. The onset of the dark baryon occurs when $\mu_n=m_\chi^*$, while chemical equilibrium requires
$\mu_n=\mu_\chi$ above the threshold.}
    \label{fig:chemical_potential}
\end{figure}
Figure-\ref{fig:chemical_potential} displays the density dependence
of the neutron chemical potential $\mu_n$, the dark-baryon chemical
potential $\mu_\chi$, and the in-medium dark-baryon mass $m_\chi^*$.
As discussed in Sec. \ref{DM-Model}, the dark state first becomes energetically accessible when $\mu_n=m_\chi^*$ at vanishing dark-baryon Fermi
momentum. Since the Higgs-induced shift of $m_\chi^*$ is negligible for the adopted parameters, the threshold is numerically indistinguishable from the condition
$\mu_n=m_\chi$.
Above this threshold, the
equilibrium abundance of $\chi$ is determined from $\mu_n=\mu_\chi$.
The onset density exhibits a clear dependence on $m_\chi$. Increasing
the dark-baryon mass from $1250$ to $1400$ MeV requires a larger
neutron chemical potential and consequently shifts the appearance of
$\chi$ toward higher density. The $m_\chi=1250$ MeV configuration
therefore develops the earliest onset and the largest dark-baryon
population, whereas the influence of $\chi$ is progressively reduced
for $m_\chi=1300$ and $1400$ MeV.

The hadronic composition also has an important influence on the
conversion threshold. In nucleonic matter, the neutron chemical
potential rises relatively rapidly with density, and the onset of
$\chi$ for $m_\chi=1250$ MeV occurs at approximately
$n_b\simeq2.9~n_0$, as inferred from the present composition results.
When hyperons and $\Delta$ resonances are included, the total baryon
number is redistributed among several Fermi seas. This reduces the
rate at which $\mu_n$ increases with density and shifts the dark-baryon
threshold to approximately
$n_b\simeq 3.5~n_0$ for the same benchmark mass.

The shift of the threshold provides direct evidence of the competition
between the dark and conventional non-nucleonic sectors. The hyperons
and $\Delta$ resonances do not arise through a sequential decay of
neutrons. Instead, all particle populations are determined
simultaneously from chemical equilibrium, charge neutrality, and total
baryon-number conservation. Their appearance reduces the chemical
driving force available for producing $\chi$, thereby delaying and
suppressing the dark component.

\subsection{Equation of state}
\label{subsec:eos}

\begin{figure}[H]
    \centering
    \includegraphics[width=1.0\linewidth]{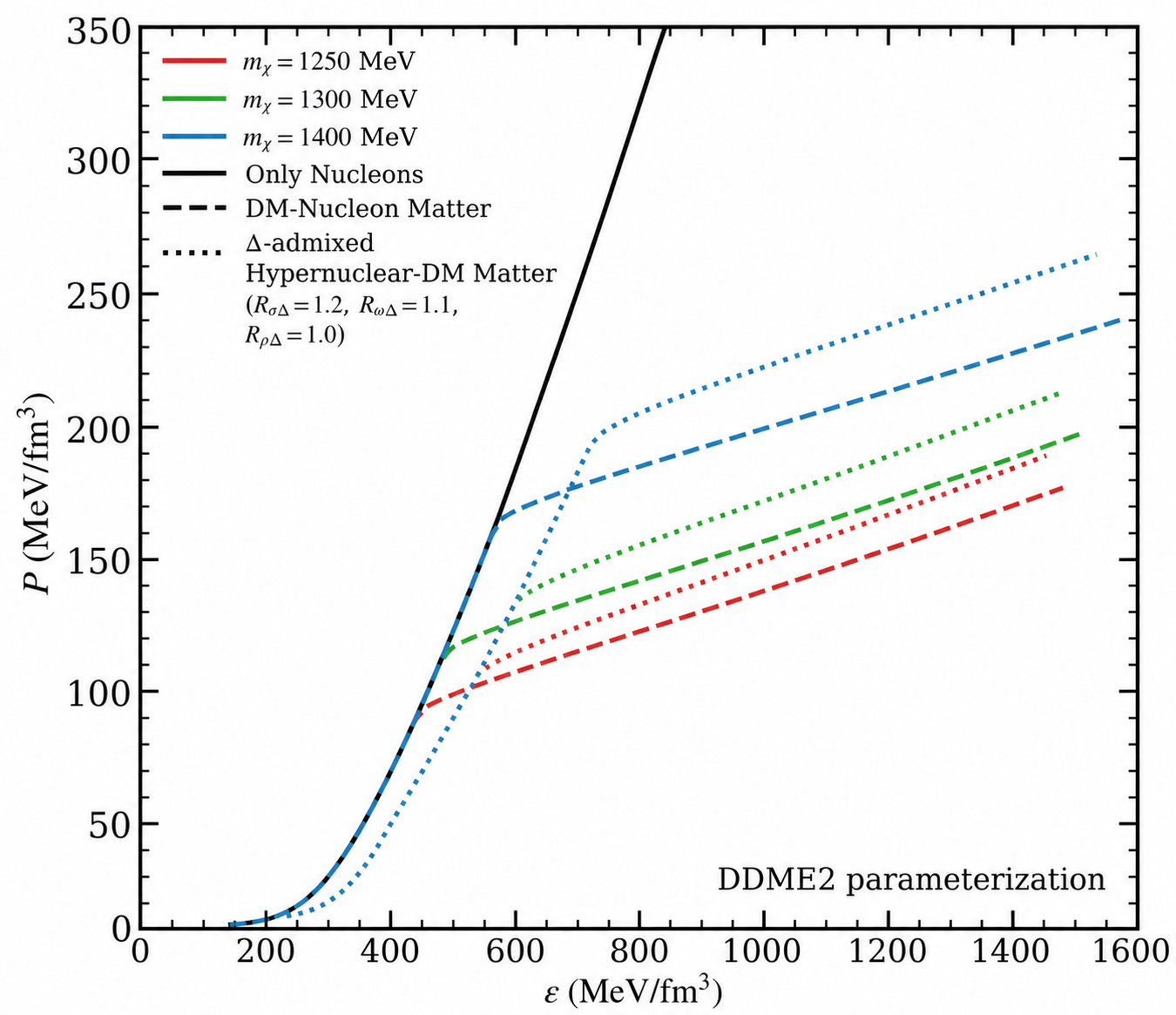}
    \caption{Pressure $P$ as a function of energy density $\varepsilon$ for purely nucleonic matter, nucleonic matter with a density-generated dark baryon, and the complete $N+Y+\Delta+\chi$ composition. Results are shown for $m_\chi=1250$, $1300$, and $1400$ MeV using the DDME2 parameterization with $R_{\sigma\Delta}=1.2$, $R_{\omega\Delta}=1.1$ and $R_{\rho\Delta}=1.0$.}
    \label{fig:EOS}
\end{figure}

The resulting EOSs are shown in
Fig.~\ref{fig:EOS}. We compare purely nucleonic matter, nucleonic
matter containing the density-generated dark baryon and the complete
$N+Y+\Delta+\chi$ composition. The EOS begins to deviate from the
purely nucleonic result once the first additional degree of freedom
becomes populated.

\textcolor{black}{The mass dependence of the dark-baryon abundance follows from Eq.\eqref{22a}, valid above the population threshold. At fixed \(\mu_n\), a smaller \(m_\chi^*\simeq m_\chi\) produces a larger $\chi$ density. In the self-consistent solutions, the neutron chemical potential also responds to the changing composition. Figure-\ref{fig:EOS} shows that the lighter-\(m_\chi\) benchmarks produce a greater reduction in pressure at a given energy density: baryon number is redistributed from the strongly interacting hadronic sector into the weakly interacting $\chi$ component, reducing the pressure support despite the additional $\chi$ kinetic pressure. Increasing \(m_\chi\) reduces this effect over the density range shown. Similar EOS softening has been reported in prescribed-admixture calculations \cite{2024PhRvD.110f3001K, 3zy4-h2ty}, although the $\chi$ abundance is determined here by conversion equilibrium.}


\begin{figure*}[t]
    \centering
    \includegraphics[width=0.81\linewidth]{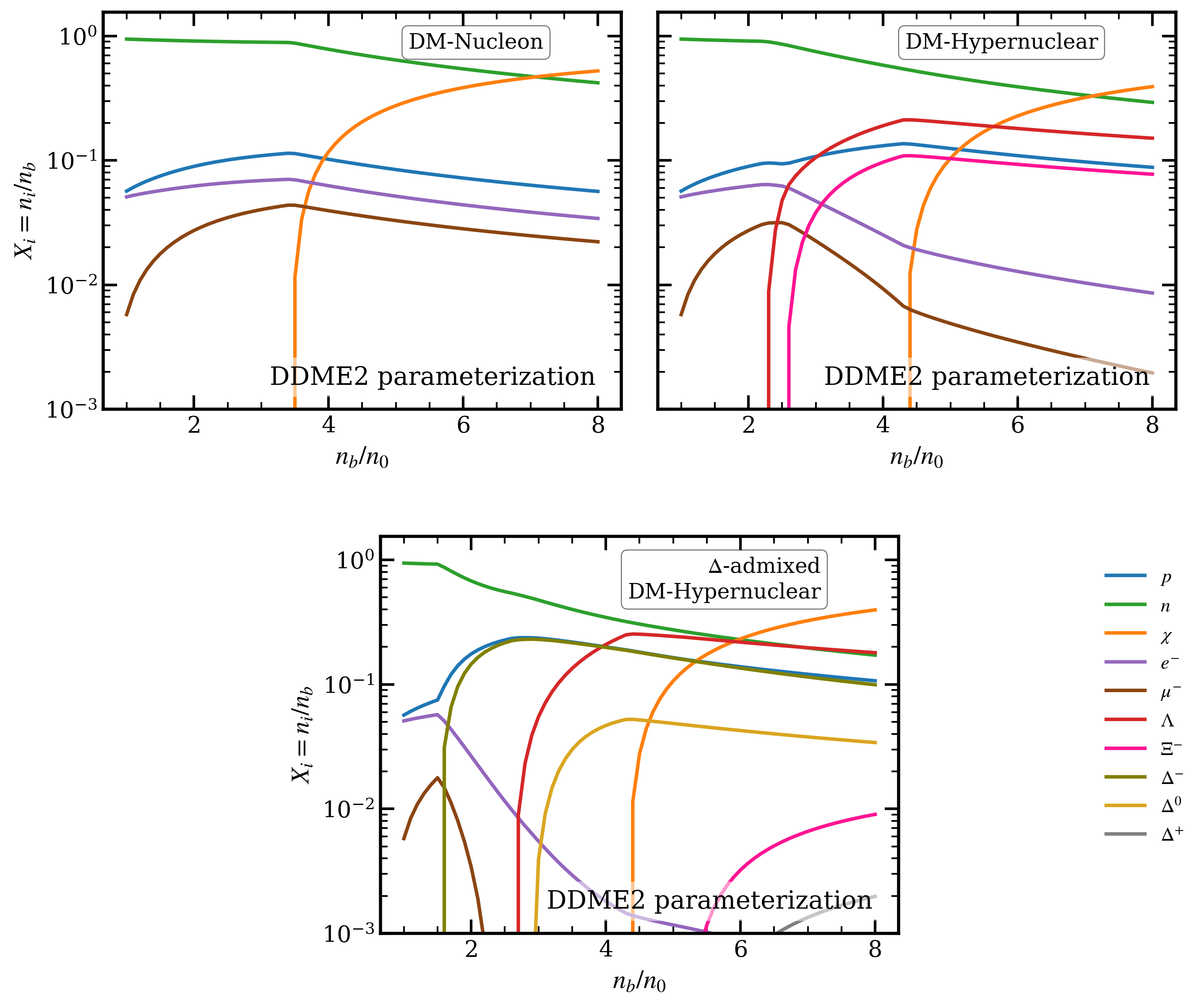}
    \caption{Particle abundances $n_i$ (in units of $n_b$) as a function of normalized baryon number density $n_b/n_0$ for the matter composition considering (i) DM-Nucleons, (ii) DM-Hypernuclear Matter and, (iii) \(\Delta\)-admixed Hypernuclear-DM Matter for DM mass \(M_{\chi}=1400\) MeV, considering DDME2 parameterization. 
    }
    \label{fig:fraction}
\end{figure*}

The complete $N+Y+\Delta+\chi$ EOS is softer than the corresponding
nucleonic+$\chi$ result. This additional reduction in pressure arises
from the appearance of hyperons and $\Delta$ resonances, which provide
further channels for redistributing the total baryon number. At the
same time, these hadronic species reduce the neutron chemical
potential and delay the appearance of $\chi$. The full EOS therefore
reflects two competing effects: direct softening due to additional
hadronic degrees of freedom and partial suppression of the
dark-baryon fraction. The changes in slope visible in
Fig.~\ref{fig:EOS} should consequently be interpreted together with
the particle fractions presented in Fig.~\ref{fig:fraction}.
We emphasize that this softening is not generated by the Higgs mean field. The latter gives a numerically negligible correction for the couplings considered here; the dominant effect arises from the conversion-induced redistribution of baryon number among the available fermionic degrees of freedom.

\subsection{NS structure solutions}
\label{subsec:mass_radius}

\begin{figure}[H]
    \centering
    \includegraphics[width=1.0\linewidth]{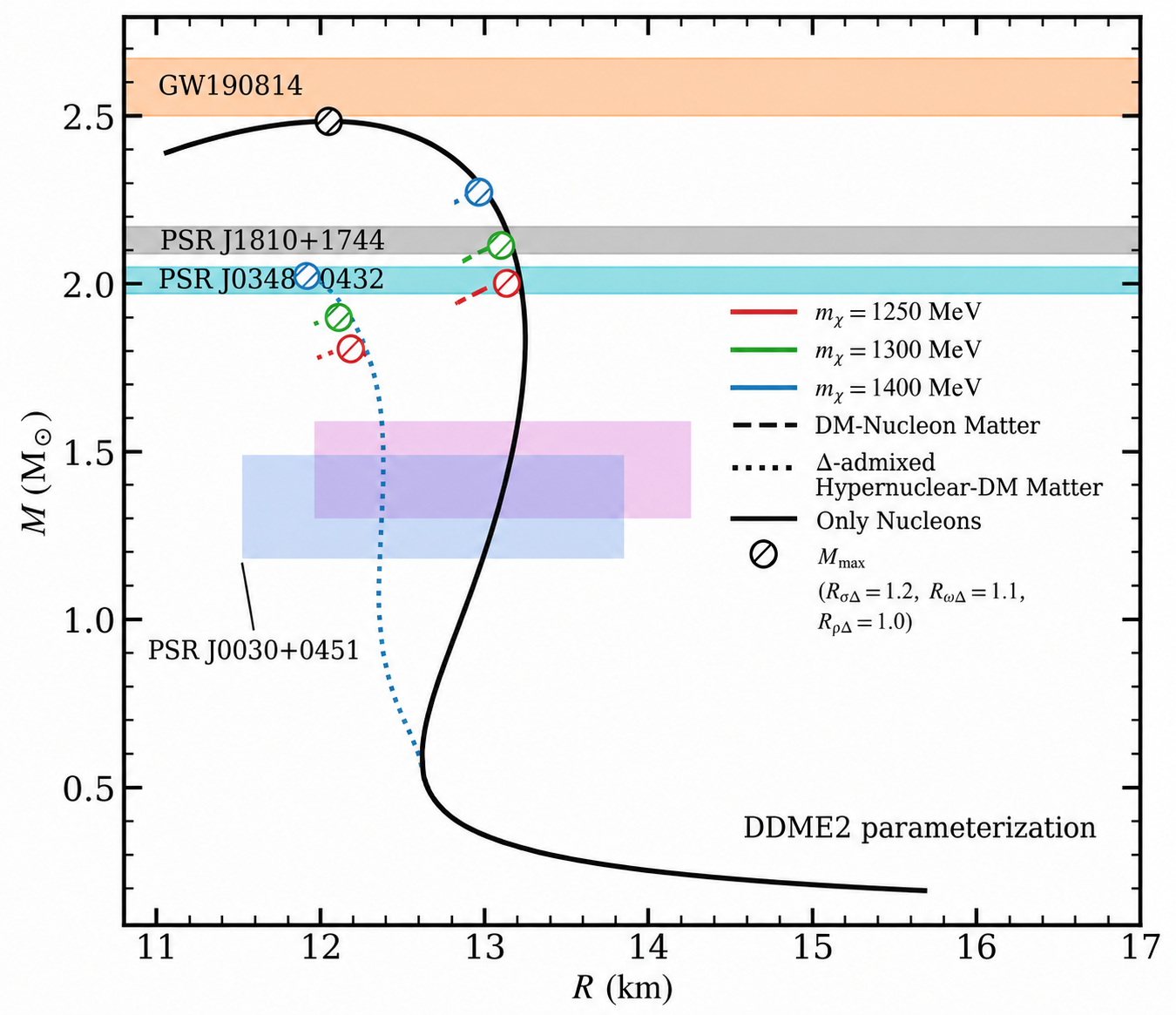}
    \caption{Gravitational mass $M$ as a function of radius $R$ for purely nucleonic matter, nucleonic matter containing $\chi$, and the complete $N+Y+\Delta+\chi$ composition. Results are shown for $m_\chi=1250$, $1300$ and $1400$ MeV. 
    The astrophysical constraints from GW190814 \cite{2020ApJ...896L..44A}, PSR J1810+1744 \cite{Romani_2021}, PSR J0030+0451 \cite{2019ApJ...887L..24M, 2019ApJ...887L..21R}, J0348 + 0432 \cite{Antoniadis_2013} are represented by the shaded regions.
    }
    \label{fig:MR}
\end{figure}

The stellar sequences corresponding to the equations of state shown
in Fig.\ref{fig:EOS} are presented in Fig.\ref{fig:MR}.
For the purely nucleonic DDME2 EOS, the maximum mass
and its corresponding radius are
$M_{\max}^{N}=2.4831\,M_{\odot}$, $R_{\max}^{N}=12.0432~{\rm km}$.
The introduction of a density-generated dark-baryon component reduces the pressure support at high density and shifts the turning point of
the stellar sequence toward lower masses. 
This effect is strongest for
the smallest value of $m_{\chi}$, for which the dark-baryon threshold
is reached at a lower density and a larger fraction of the stellar core is populated by $\chi$.

For nucleonic matter containing $\chi$, the maximum masses are
$M_{\max}^{N+\chi}
= 2.0011,\; 2.1136,\; 2.2721\,M_{\odot}$
for $m_{\chi}=1250$, $1300$ and $1400$ MeV, respectively. 
Relative to the purely nucleonic value, these results correspond to reductions
of approximately $19.4\%$, $14.9\%$, and $8.5\%$. The monotonic
increase of $M_{\max}$ with $m_{\chi}$ follows from the progressive
displacement of the dark-baryon threshold toward higher density:
a heavier $\chi$ occupies a smaller part of the stable stellar branch
and therefore produces less softening.
The radii of the corresponding maximum-mass configurations are
$R_{\max}^{N+\chi} = 13.1293,\; 13.0941,\; 12.9576~{\rm km}$.
Since these radii correspond to different gravitational masses and
central densities, their ordering should not be interpreted as a
fixed-mass measure of stiffness.
\textcolor{black}{The radius, central density and compactness at a common gravitational mass provide a more appropriate comparison.}

\begin{figure}[H]
    \centering
    \includegraphics[width=1.0\linewidth]{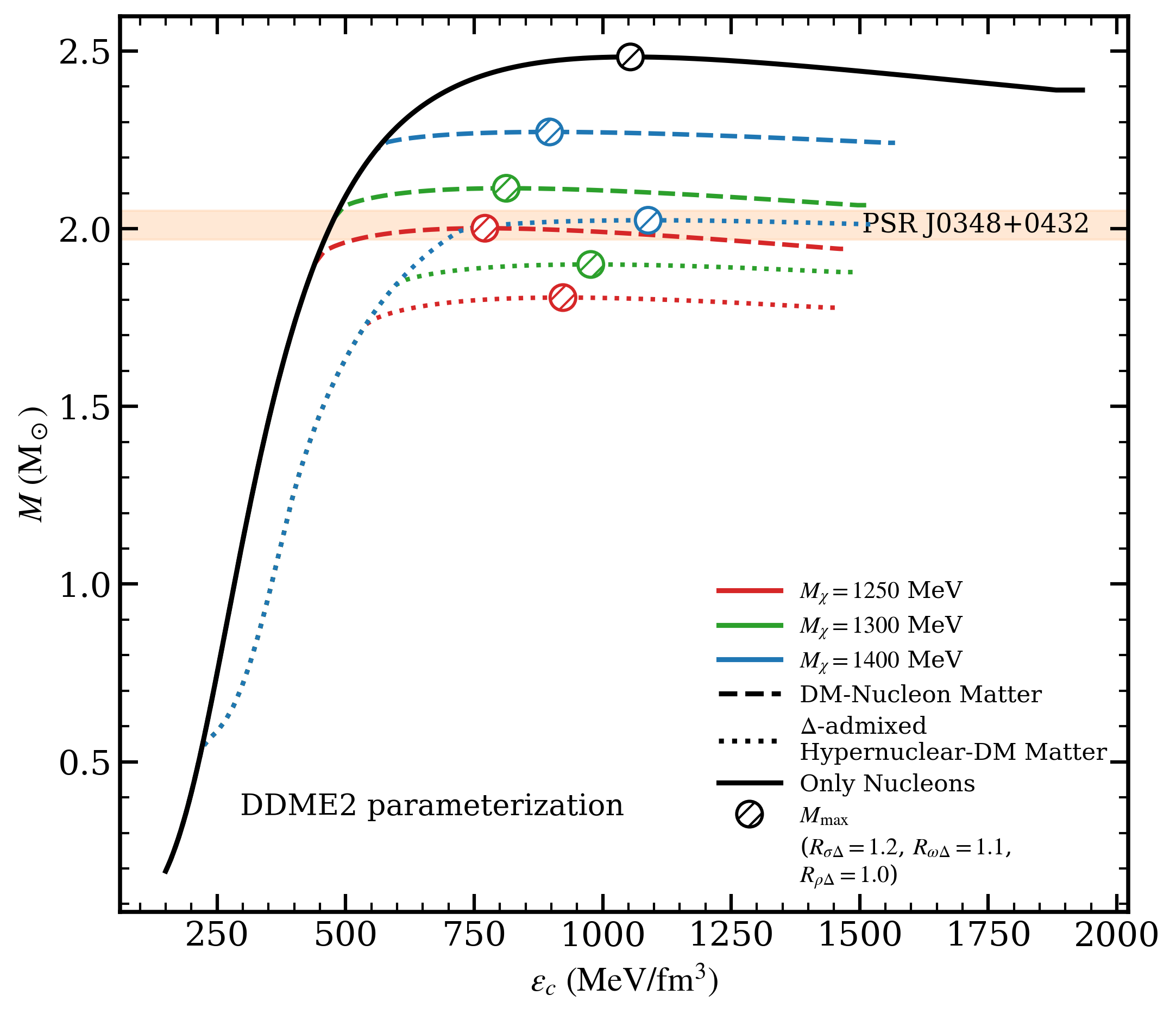}
    \caption{\textcolor{black}{Gravitational mass \(M\) as a function of central energy density \(\varepsilon_c\) for the indicated compositions. The marked points denote the maximum-mass configurations. Results are shown for $m_\chi=1250$, $1300$ and $1400$ MeV. 
    The astrophysical constraints from J0348 + 0432 \cite{Antoniadis_2013} are represented by the shaded regions.}}
    \label{fig:E_vs_M}
\end{figure}

The simultaneous inclusion of hyperons and $\Delta$ resonances
produces an additional reduction in the maximum mass. For the complete
$N+Y+\Delta+\chi$ composition, the calculated values are $M_{\max}^{N+Y+\Delta+\chi} = 1.80,\; 1.90,\;
2.02\,M_{\odot}$ for $m_{\chi}=1250$, $1300$, and $1400$ MeV, respectively. 
These values are approximately $9.7\%$, $10.1\%$, and $10.9\%$ lower than
the corresponding nucleonic+$\chi$ results. Relative to the purely
nucleonic model, the total reductions amount to approximately $27.3\%$, $23.5\%$, and $18.5\%$, respectively.
The corresponding maximum-mass radii are
$R_{\max}^{N+Y+\Delta+\chi} = 12.1784,\; 12.1064,\; 11.9079~{\rm km}$.
The $m_{\chi}=1250$ and $1300$ MeV complete-composition models do not
satisfy the conventional $2\,M_{\odot}$ requirement \cite{2010Natur.467.1081D,Arzoumanian_2018, Antoniadis_2013, 2020NatAs...4...72C}.
The $m_{\chi}=1400$ MeV case reaches
$M_{\max}=2.0236\,M_{\odot}$ and is therefore only marginally compatible with this lower mass scale. 
Within the adopted coupling scheme, the massive-pulsar constraint consequently favors a comparatively heavy dark baryon.

\textcolor{black}{Figure-\ref{fig:E_vs_M} shows the gravitational mass as a function of central energy density, \(M(\varepsilon_c)\). The marked extrema satisfy \(dM/d\varepsilon_c=0\). For the connected, nonrotating equilibrium sequences considered here, the first maximum provides the usual turning-point diagnostic when perturbations follow the same barotropic equilibrium relation. The descending branches are displayed to identify these extrema and are excluded from the stellar observables quoted for the ordinary stable branch. This diagnostic does not establish stability for perturbations with frozen composition or finite conversion timescales.}

A linear interpolation between the $m_{\chi}=1300$ and $1400$ MeV
results gives an approximate crossing of
$M_{\max}=2\,M_{\odot}$ near $m_{\chi}\simeq1.38~{\rm GeV}$ for $R_{\sigma\Delta}=1.2$. This value should be regarded only as an
indicative scale rather than a formal lower limit, 
because a finer scan in $m_\chi$, together with the uncertainty in the $\Delta$-meson couplings and in the assumed conversion-equilibrium limit, has not yet been incorporated.
Accordingly, the preference for larger $m_\chi$ should be understood as conditional on the assumed $(n\leftrightarrow\chi)$ chemical-equilibrium limit and the adopted hadronic interaction scheme, rather than as a model-independent constraint on the dark-baryon mass.
\textcolor{black}{At the common gravitational mass $M=1.8M_\odot$, all three nucleonic+$\chi$ configurations lie below their respective dark-baryon onset masses, which are $1.936$, $2.063$, and $2.243M_\odot$ for
$m_\chi=1250$, $1300$, and $1400$ MeV.
Consequently, these configurations contain no $\chi$ population and coincide with the purely nucleonic solution, giving $R_{1.8}=13.2439$ km and
$n_c^{1.8}/n_0=2.6750$ within the quoted precision.
The fixed-mass comparison therefore shows that conversion has not yet modified the nucleonic configurations at this mass.}

The observational regions displayed in Fig.\ref{fig:MR} include the adopted NICER mass--radius inference for
PSR J0030+0451 \cite{2019ApJ...887L..24M, 2019ApJ...887L..21R}, PSR J0348+0432 \cite{Antoniadis_2013} and the selected massive-pulsar constraint, PSR J1810+1744 \cite{Romani_2021}. 
The GW190814 band \cite{2020ApJ...896L..44A} is shown only to illustrate the hypothesis that its secondary component may have been a NS; it should not be interpreted as a model-independent upper bound on the maximum mass of nonrotating NSs.

\begin{table}[htbp]
\centering
\caption{Onset number densities and corresponding stellar onset masses of particles in the NS with composition of DM-Nucleon ($N+\chi$), DM-hypernuclear ($N+Y+\chi$) and $\Delta$-admixed hypernuclear-DM ($N+Y+\Delta+\chi$) compositions, for DM masses : $m_\chi=1250, 1300$ and $1400$ MeV.}
\label{tab:onset}
\resizebox{0.38\textwidth}{!}{%
\begin{tabular}{c|c|c|c|c}
\hline
\hline
$m_\chi$ & Composition & Species &  $n_i^{\rm onset}$ &  $M_i^{\rm onset}$\\
(MeV) &  &  &  $(n_0)$ & ($M_\odot$)\\
\hline

    & $N+\chi$            &   $\chi$      & 2.900 & 1.936  \\
\cline{2-5}
    &                                    & $\Lambda$   & 2.300 & 1.443  \\
    &  $N+Y+\chi$             & $\Xi^-$     & 2.600 & 1.640 \\
    &                                    & $\chi$      & 3.200 & 1.861  \\
\cline{2-5}
1250    &                                    & $\Delta^-$  & 1.600 & 0.518 \\
    &                                    & $\Delta^0$  & 2.800 & 1.372 \\
    &   $N+Y+\Delta+\chi$     & $\Lambda$   & 2.700 & 1.293  \\
    &                                    & $\chi$      & 3.500 & 1.735 \\
\hline

    & $N+\chi$            &   $\chi$      & 3.100 & 2.063\\
\cline{2-5}
    &                                   & $\Lambda$   & 2.300 & 1.443  \\
    &   $N+Y+\chi$           & $\Xi^-$     & 2.600 & 1.640 \\
    &                                    & $\chi$      & 3.600 & 1.953 \\
\cline{2-5}
1300    &                                    & $\Delta^-$  & 1.600 & 0.518 \\
    &                                    & $\Delta^0$  & 2.800 & 1.372 \\
    &   $N+Y+\Delta+\chi$     & $\Lambda$   & 2.700 & 1.293 \\
    &                                    & $\chi$      & 3.800 & 1.845 \\
\hline

    & $N+\chi$            &   $\chi$      & 3.500 & 2.243 \\
\cline{2-5}
    &                                    & $\Lambda$   & 2.300 & 1.443 \\
    &   $N+Y+\chi$           & $\Xi^-$     & 2.600 & 1.640 \\
    &                                    & $\chi$      & 4.400 & 2.060 \\
\cline{2-5}
1400    &                                    & $\Delta^-$  & 1.600 & 0.518  \\
    &                                    & $\Delta^0$  & 2.800 & 1.372 \\
    &                                    & $\Delta^+$  & 5.299 & 2.014 \\
    &   $N+Y+\Delta+\chi$     & $\Lambda$   & 2.700 & 1.293 \\
    &                                    & $\Xi^-$     & 5.399 & 2.015 \\
    &                                    & $\chi$      & 4.400 & 1.994 \\
\hline
\hline
\end{tabular}}
\end{table}

\subsection{Equilibrium particle composition}
\label{subsec:composition}

Figure-\ref{fig:fraction} compares the density dependence of the particle populations for nucleonic matter containing $\chi$,
hypernuclear matter containing $\chi$, and
$\Delta$-admixed hypernuclear matter containing $\chi$, for $m_{\chi}=1400$ MeV. 
The plotted quantity is the local particle fraction $X_i=n_i/n_b$.

In nucleonic+$\chi$ matter, the dark-baryon density becomes nonzero at approximately $n_b\simeq3.5n_0$ and subsequently increases rapidly. 
Its growth is accompanied primarily by a reduction of the neutron density, reflecting the chemical-equilibrium condition $\mu_n=\mu_{\chi}$ and total baryon-number conservation. 
The proton, electron and muon densities exhibit comparatively modest changes, since they remain constrained by charge neutrality and
$\beta$ equilibrium.

A summary of the onset densities of various exotic particle spectrum in dense matter is provided in Table-\ref{tab:onset}.
When hyperons are admitted, the $\Lambda$ and $\Xi^{-}$ states become
populated before the dark baryon. Their appearance redistributes the total baryon density and modifies both the neutron and electron
chemical potentials. As a consequence, the onset of $\chi$ is shifted
to a higher density than in nucleonic+$\chi$ matter, and its abundance at a given $n_b$ is reduced. 
The negatively charged $\Xi^{-}$ also
partially replaces the leptons in maintaining charge neutrality.
The composition changes further when the complete $\Delta$ quartet is included. 
The $\Delta^{-}$ appears at comparatively low density owing to its negative electric charge, followed by the $\Delta^{0}$ and,
at higher density, the $\Delta^{+}$. The early population of
$\Delta^{-}$ reduces the electron and muon densities and modifies the isovector mean field. 
The dark baryon appears only after this substantial rearrangement of the hadronic composition. 
The ordering of the thresholds therefore demonstrates that the formation of $\chi$ cannot be treated independently of the conventional non-nucleonic
degrees of freedom.

\subsection{Binary tidal response}
\label{subsec:tidal}

\begin{figure}[htbp]
    \centering
    \includegraphics[width=1.0\linewidth]{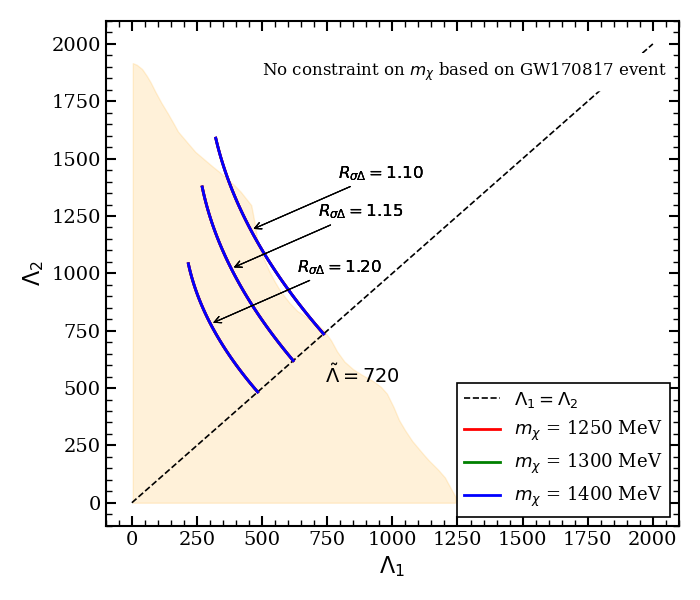}
    \caption{Tidal deformabilities $\Lambda_1$ and $\Lambda_2$ of the two components of a GW170817-like binary with chirp mass $\mathcal{M}=1.186\,M_{\odot}$. Results are shown for $R_{\sigma\Delta}=1.10$, $1.15$, $1.20$ and for the three representative dark-baryon masses considered in this work. The diagonal line corresponds to $\Lambda_1=\Lambda_2$. The shaded region represents the adopted observationally allowed region in the $\Lambda_1$--$\Lambda_2$ plane for GW170817 \cite{PhysRevX.9.011001}. The tidal response in this mass range is predominantly sensitive to the $\sigma-\Delta$ interaction and only weakly sensitive to $m_\chi$.}
    \label{fig:Tidal}
\end{figure}

Figure-\ref{fig:Tidal} shows the individual tidal deformabilities $\Lambda_1$ and $\Lambda_2$ for binary configurations with chirp mass
$\mathcal{M}=1.186\,M_{\odot}$,
corresponding to GW170817 \cite{LIGO_Virgo2017c}. 
The scalar $\Delta$-meson coupling is varied
over $R_{\sigma\Delta}=1.10$, $1.15$ and $1.20$, while the three benchmark dark-baryon masses are considered.

The trajectories obtained for $m_{\chi}=1250$, $1300$ and
$1400$ MeV are nearly indistinguishable for a fixed value of
$R_{\sigma\Delta}$. 
The near-overlap of the displayed trajectories shows that these tidal observables have little sensitivity to the three selected dark-baryon masses over the sampled component-mass interval. 
Establishing whether either component contains $\chi$ requires comparing its central density with the corresponding onset density; the absence of $\chi$ in a canonical \(1.4\,M_\odot\) star alone is insufficient to establish this for the entire binary sequence.
\vt{A more visible dependence is produced by
$R_{\sigma\Delta}$ in line with Refs.\cite{Thapa_2021, Thapa_2021_Bary}. Increasing the scalar $\Delta$ coupling enhances
the attractive interaction in the $\Delta$ sector, softens the
intermediate-density EOS, and reduces the corresponding
stellar radii and tidal deformabilities. 
The binary tidal response is therefore more sensitive to the $\Delta$ interaction than to $m_{\chi}$ for the mass range relevant to GW170817.}

\subsection{Sound speed and adiabatic index}
\label{subsec:sound_gamma}

\begin{figure}[h]
    \centering
    \includegraphics[width=0.8\linewidth]{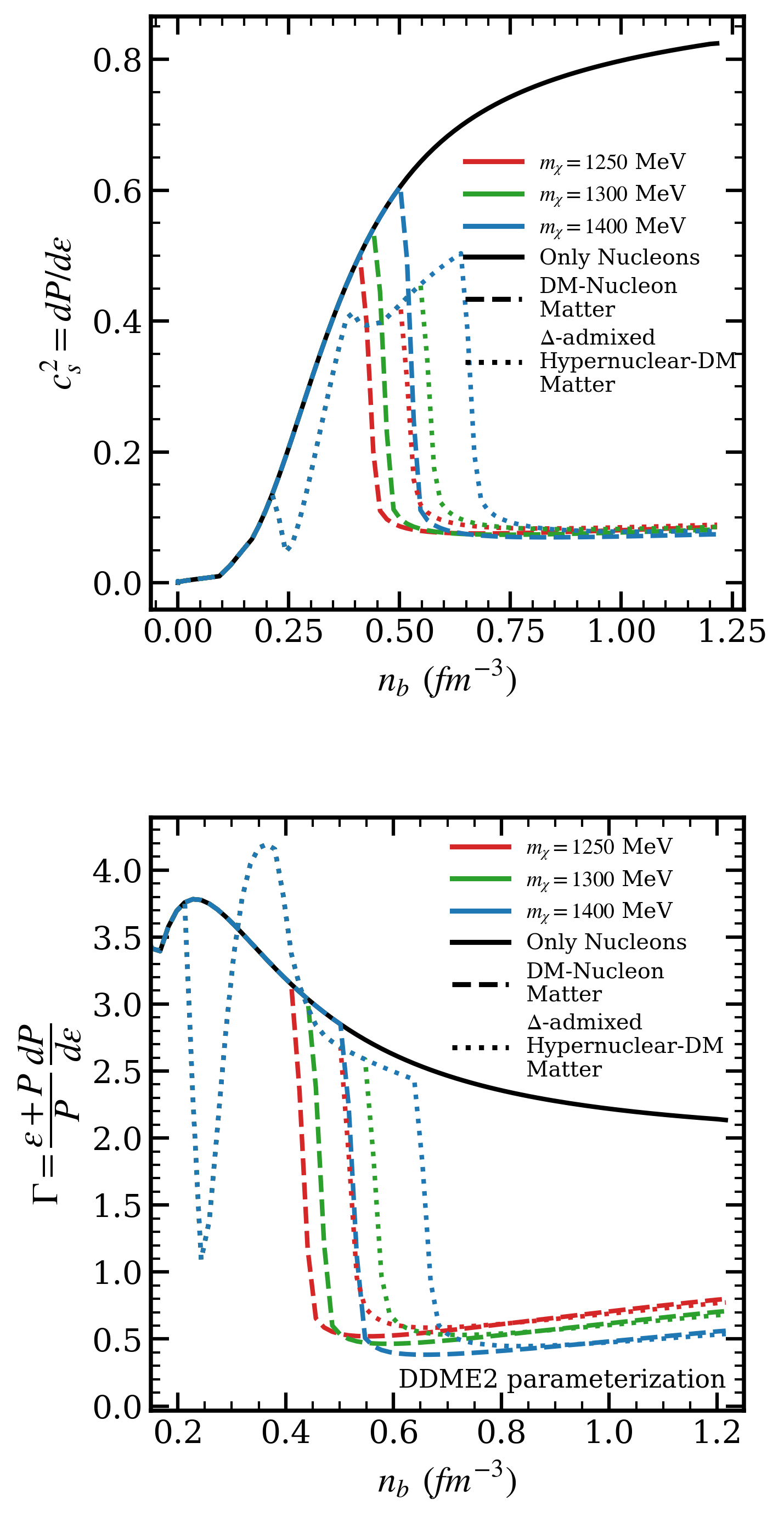}
    \caption{Squared equilibrium sound speed $c_s^2$ (top) and relativistic adiabatic index $\Gamma$ (bottom) as functions of the baryon number density $n_b$ for purely nucleonic matter, nucleonic+$\chi$ matter, and the complete $N+Y+\Delta+\chi$ composition. Results are shown for $m_\chi=1250$, $1300$, and $1400$ MeV within the DDME2 parameterization. The structures in $c_s^2$ and $\Gamma$ reflect changes in the equilibrium composition associated with the onset of non-nucleonic degrees of freedom.}
    \label{fig:sound}
\end{figure}

The squared equilibrium sound speed and the relativistic adiabatic
index are displayed in Fig.\ref{fig:sound}. 
They are defined as \cite{2022ApJ...939L..34A}
\begin{equation}
c_s^2=\frac{dP}{d\varepsilon},
\qquad
\Gamma=
\frac{\varepsilon+P}{P}
\frac{dP}{d\varepsilon}.
\end{equation}
The derivative $(dP/d\epsilon)$ is evaluated numerically from the tabulated equilibrium EOS, using the neighboring EOS points along the density sequence. The resulting structures in $(c_s^2)$ and $(\Gamma)$ are associated with changes in the equilibrium particle composition at the corresponding threshold densities.
For the purely nucleonic EOS, the squared equilibrium sound speed
$c_s^2=dP/d\varepsilon$ increases smoothly with density, reflecting
the progressive stiffening of the nucleonic medium. In contrast, the
appearance of the dark baryon produces a pronounced reduction in
$c_s^2$. This feature originates from the rearrangement of the
equilibrium composition at the $\chi$ threshold: once the additional
degree of freedom becomes populated, the pressure increases more
slowly with energy density, resulting in a locally softer EOS. The
location of this structure shifts systematically toward higher
density as $m_\chi$ is increased from $1250$ to $1400$ MeV, in
accordance with the corresponding displacement of the dark-baryon onset discussed in Sec.~\ref{subsec:chemical_potential}.
And for the complete $\Delta$-admixed hypernuclear composition,
additional structures appear in $c_s^2$ before the principal
dark-baryon-induced reduction. These features are associated with
successive changes in the equilibrium particle content, particularly
the appearance of $\Delta$ resonances and hyperons. 
Their population modifies the scalar and vector mean fields and redistributes the total
baryon density among the available fermionic species. 
This behaviour of \(\Delta\)-resonances is thoroughly studied in Refs.\cite{Thapa_2021, Parmar_2025}.

\begin{figure*}
    \centering
    \includegraphics[width=0.8\linewidth]{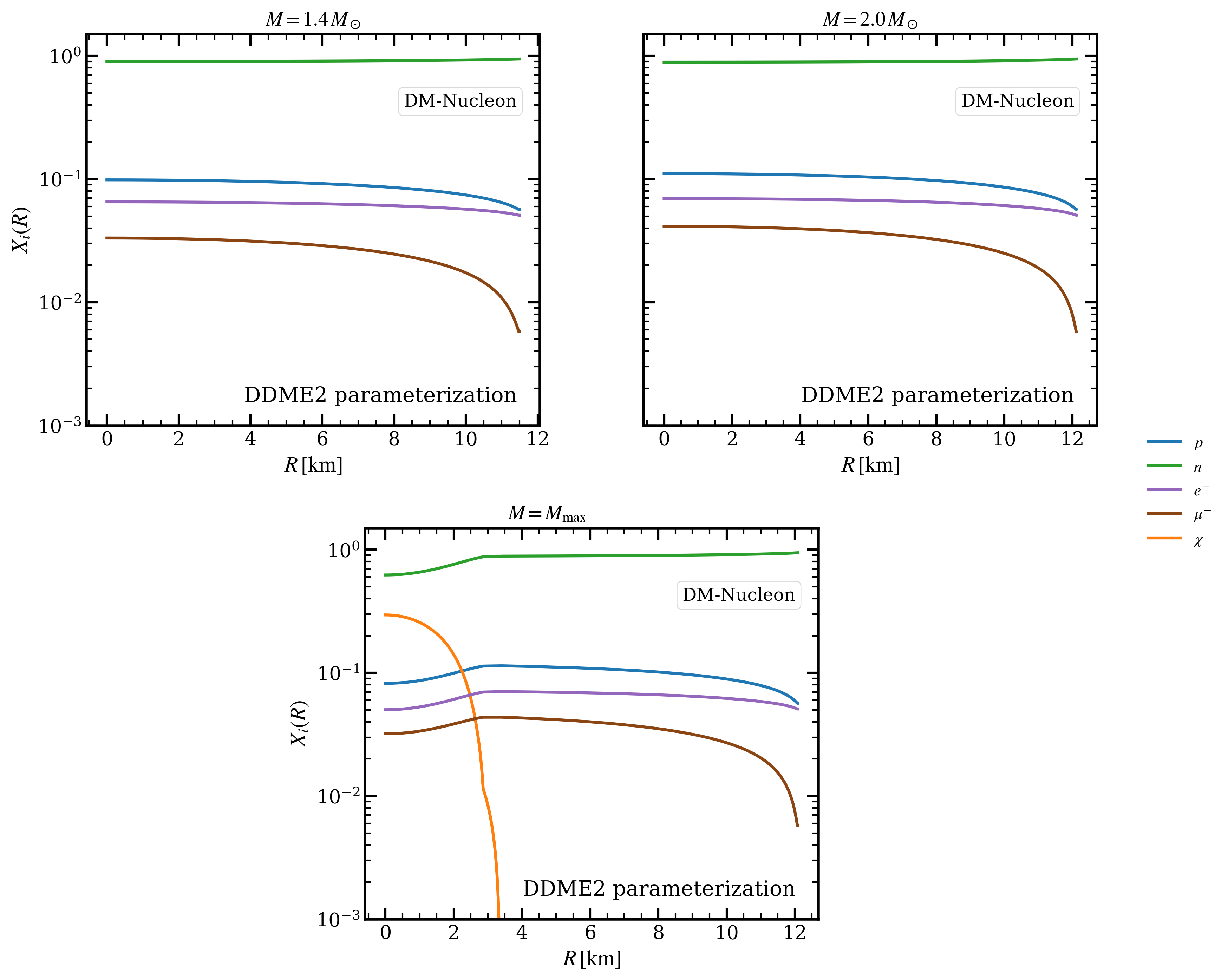}

    \vspace{2 mm}

    \centering
    \includegraphics[width=0.8\linewidth]{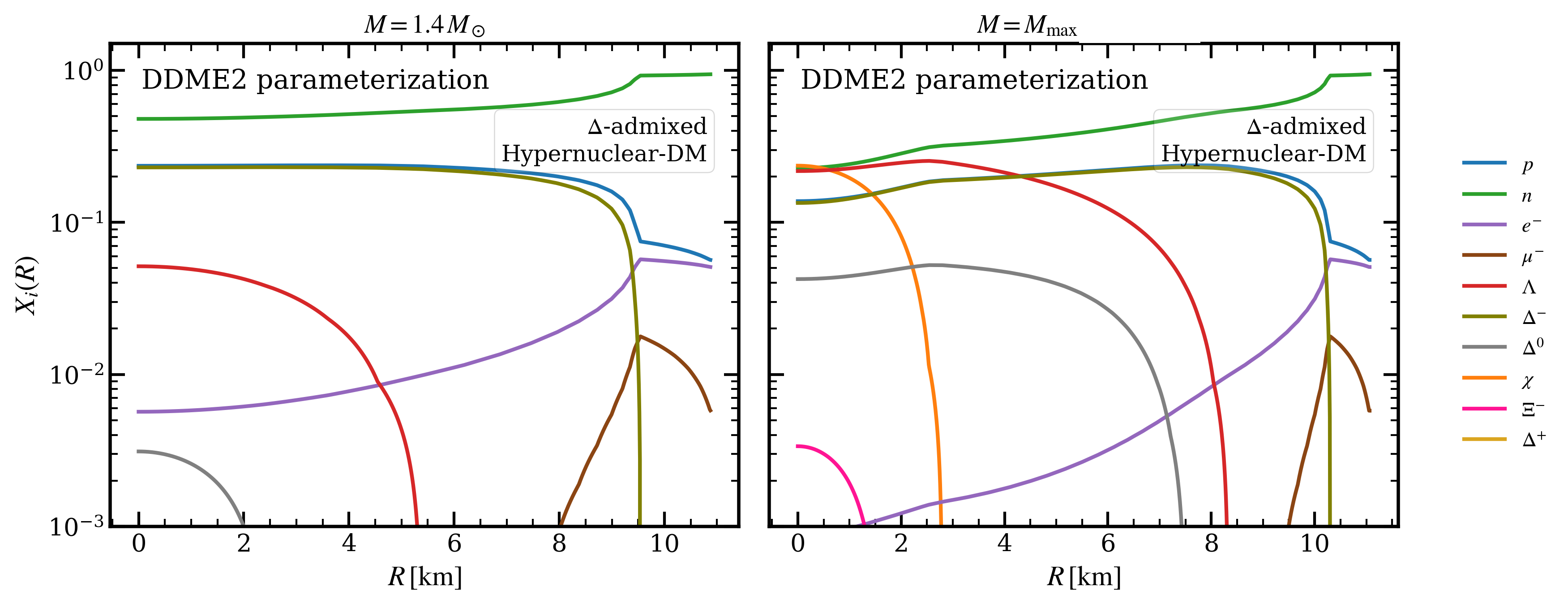}
    \caption{Radial particle-abundance profiles for $m_\chi=1400$ MeV. The upper panels correspond to nucleonic+$\chi$ matter for $M=1.4\,M_\odot$, $2.0\,M_\odot$, and the maximum-mass configuration. The lower panels show the full-composition profiles for \(M=1.4\,M_\odot\) and \(M=M_{\max}\). The dark baryon is absent from the $1.4~M_\odot$ and $2~M_\odot$ nucleonic$+\chi$ configurations and is confined to the inner few kilometres of the most massive stars. The ordinate denotes the local particle fraction $X_i(r)=n_i(r)/n_b(r)$.}
    \label{fig:Fraction_vs_Radius}
\end{figure*}

As a result, the evolution of the neutron chemical potential is altered and the
onset of $\chi$ is shifted toward higher density. The displacement of
the corresponding structure in $c_s^2$ is therefore consistent with
the chemical-potential behaviour shown in Fig.~\ref{fig:chemical_potential} and with the
particle thresholds summarized in Fig.~\ref{fig:fraction} and Table~\ref{tab:onset}.
The bottom panel of Fig.~\ref{fig:sound} shows the equilibrium adiabatic index, \(\Gamma_{\mathrm{eq}}=(\varepsilon+P)P^{-1}dP/d\varepsilon\). This quantity describes compression along the chemical-equilibrium sequence and need not equal the adiabatic index governing oscillations when particle fractions cannot equilibrate within an oscillation period.
A comparatively large value of
$\Gamma$ corresponds to a stronger resistance against compression, whereas a reduction in $\Gamma$ indicates an increase in
compressibility and hence a locally softer response of the EOS. 
The appearance of $\chi$ produces a pronounced decrease in $\Gamma$,
consistent with the behaviour observed in $c_s^2$. Increasing $m_\chi$ shifts this feature toward higher density because the formation of the dark baryon is progressively delayed. 
For the complete $N+Y+\Delta+\chi$ composition, the sharp reduction occurs
at a still higher density, since the prior appearance of hyperons and $\Delta$ resonances modifies the neutron chemical potential and delays the dark-baryon threshold.
\textcolor{black}{Over the density interval used for the reported configurations, the equilibrium EOS satisfies \(0<dP/d\varepsilon<1\), corresponding to positive equilibrium compressibility and a subluminal equilibrium sound speed. These conditions alone do not establish dynamical radial stability; the equilibrium turning-point diagnostic is shown in Fig. \ref{fig:E_vs_M}.}


\subsection{Radial composition of representative stellar configurations}
\label{subsec:radial_composition}

\vt{Figure-\ref{fig:Fraction_vs_Radius} displays the radial distributions of
the particle abundances for $m_\chi=1400$ MeV. The upper panels
correspond to nucleonic matter in chemical equilibrium with the dark
baryon, whereas the lower panels show the complete
$\Delta$-admixed hypernuclear composition. The radial profiles provide
a direct measure of the spatial extent of the dark component and of
its coexistence with conventional non-nucleonic degrees of freedom.
For the nucleonic+$\chi$ composition, neither the
$1.4\,M_\odot$ nor the $2.0\,M_\odot$ configuration contains a
visible dark-baryon component. Their central densities therefore
remain below the threshold required for the production of
$m_\chi=1400$ MeV dark baryons. The proton, electron and muon abundances vary smoothly with radius, while neutrons remain the dominant constituent throughout the stellar interior.

A qualitatively different structure is obtained for the maximum-mass nucleonic+$\chi$ configuration. 
In this case, the central density
exceeds the dark-baryon threshold and a compact $\chi$-rich region develops in the inner core. The dark abundance is largest at the
stellar center and decreases rapidly with increasing radius, becoming negligible at approximately $R\simeq2$--$3$ km. 
Outside this region, the star is composed almost entirely of ordinary nucleonic matter.
Thus, for the relatively heavy dark baryon considered here, the conversion process does not generate an extended dark component but
rather a spatially localized central core.
The lower panels demonstrate that the internal composition is substantially rearranged when hyperons and $\Delta$ resonances are included. 
The $1.4\,M_\odot$ configuration contains non-nucleonic
hadronic species in its inner region, although no appreciable $\chi$ component is present. The occurrence of heavy baryons in a
canonical-mass star therefore precedes the formation of the dark baryon for the adopted value of $m_\chi$.
For the maximum-mass $\Delta$-admixed hypernuclear configuration,
the central region contains $\chi$, hyperons, and several members of
the $\Delta$ quartet simultaneously. The dark component is again
confined to the innermost few kilometres, while the hyperonic and
$\Delta$ populations extend over a larger fraction of the stellar
core. In particular, the $\Lambda$ and negatively charged baryonic
species remain populated at radii well beyond the boundary of the
dark core. The resulting radial hierarchy may therefore be summarized
schematically as $R_\chi < R_Y,\;R_\Delta < R_{\rm star}$, where $R_\chi$ denotes the radius of the region containing a non-vanishing dark-baryon abundance.

The radial profiles further show that canonical-mass observables need not be directly sensitive to $\chi$ even when the dark baryon has a
significant effect on the maximum-mass sequence. For
$m_\chi=1400$ MeV, the dark component is restricted to the central region of the most massive configurations, whereas the $1.4\,M_\odot$ star remains dark-baryon free. 
This explains the weak dependence of $R_{1.4}$ and $\Lambda_{1.4}$ on $m_\chi$ in the heavy
dark-baryon regime (refer to Fig.\ref{fig:Tidal}).
}
Table \ref{tab:summary} quantifies the dependence of the maximum-mass
configuration and the central dark-baryon fraction on
$m_\chi$. Increasing $m_\chi$ delays the onset of $\chi$
and reduces the central dark-baryon fraction of the maximum-mass configuration, $X_{\chi,c}^{\max}$, thereby recovering part
of the high-density pressure support. The inclusion of
hyperons and $\Delta$ resonances further reduces the
maximum mass and suppresses the central dark-baryon
fraction through their competition with $\chi$.

\subsection{Dependence on the dark-baryon mass and $\Delta$ coupling}
\label{subsec:parameter_dependence}

\begin{figure*}
    \vspace*{-1cm}
    \includegraphics[width=0.85\linewidth]{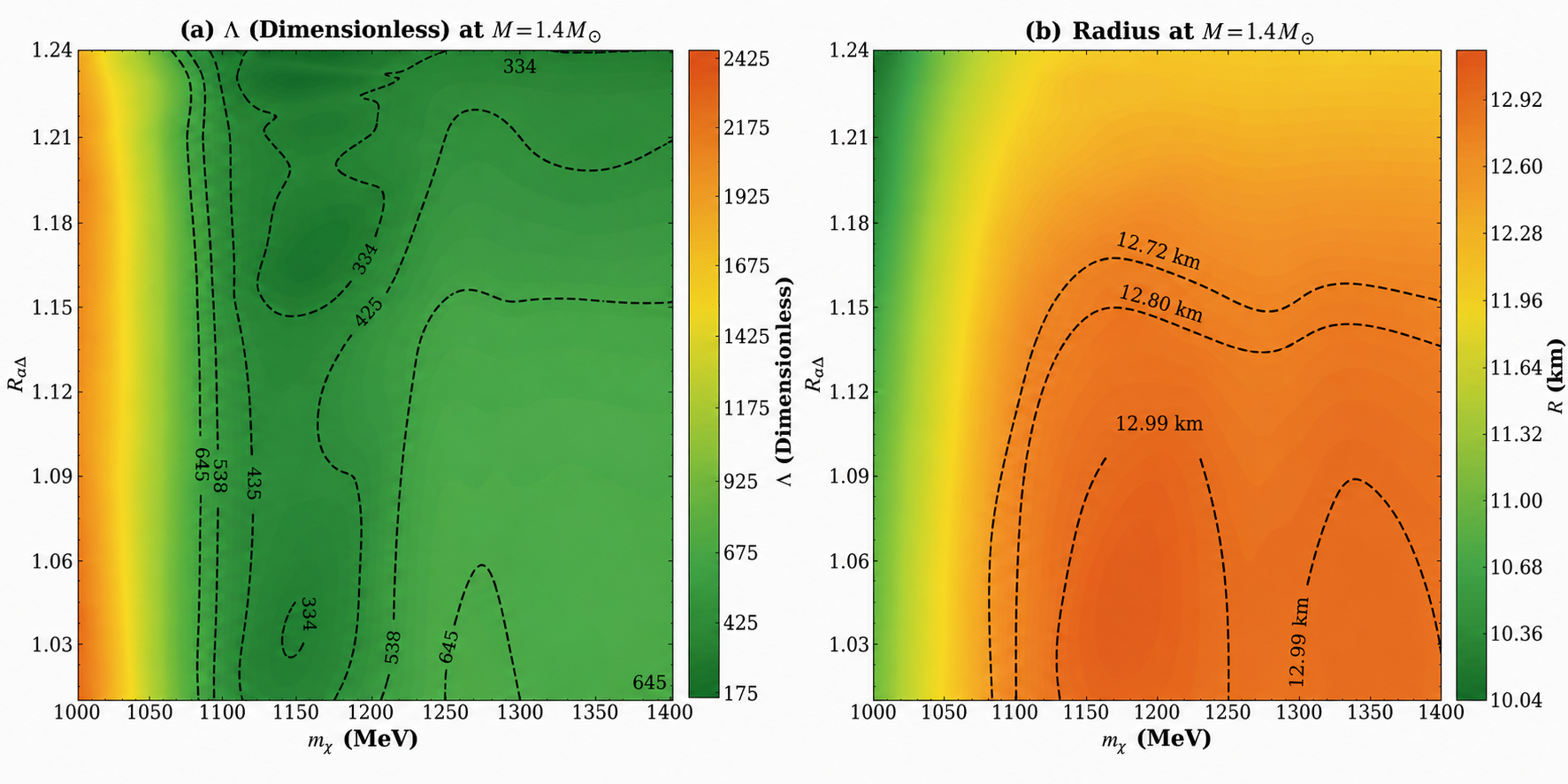}
    \label{fig:Contour1}
    \includegraphics[width=0.51\linewidth]{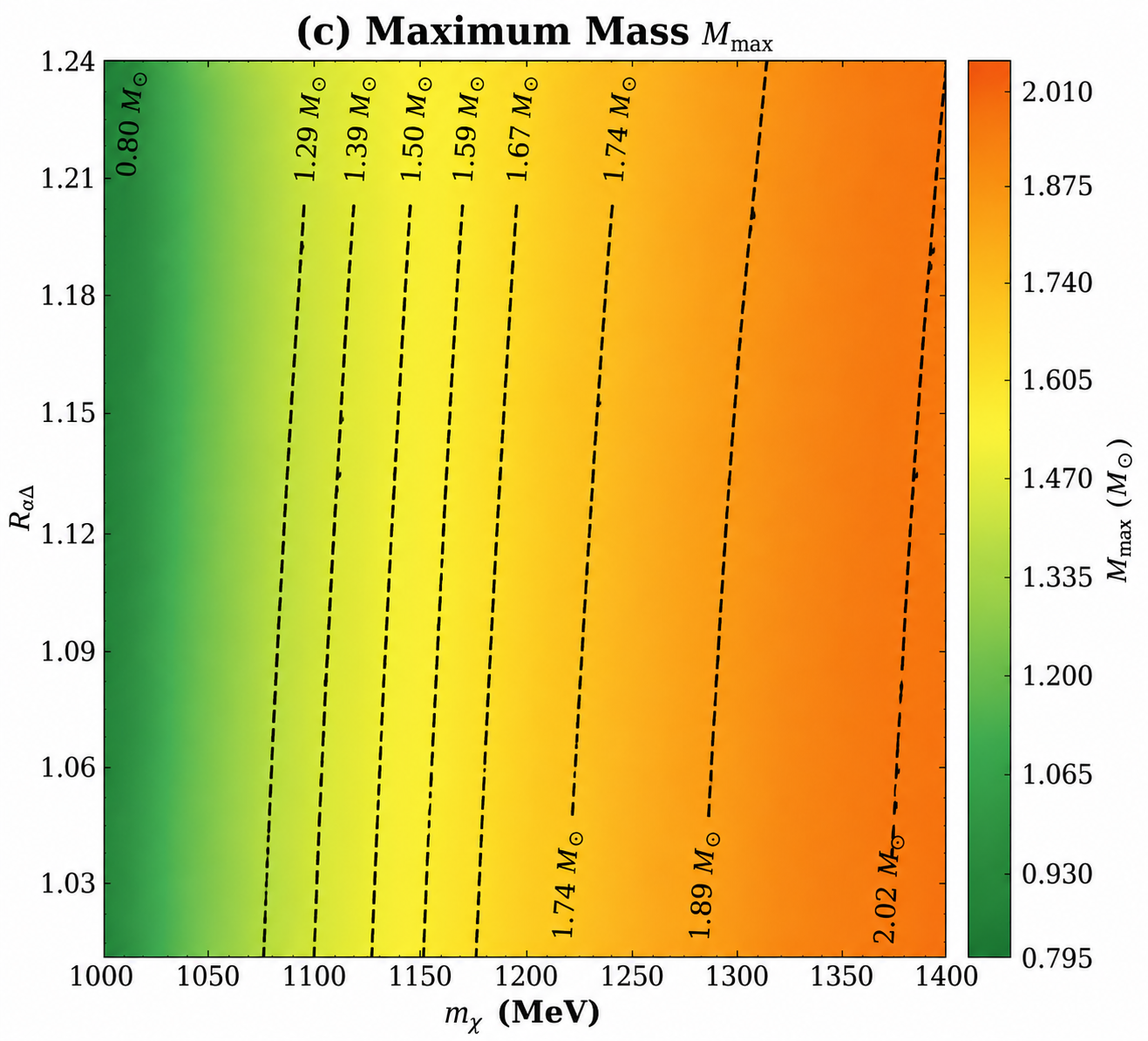}
    \label{fig:Contour2}
    \includegraphics[width=0.9\linewidth]{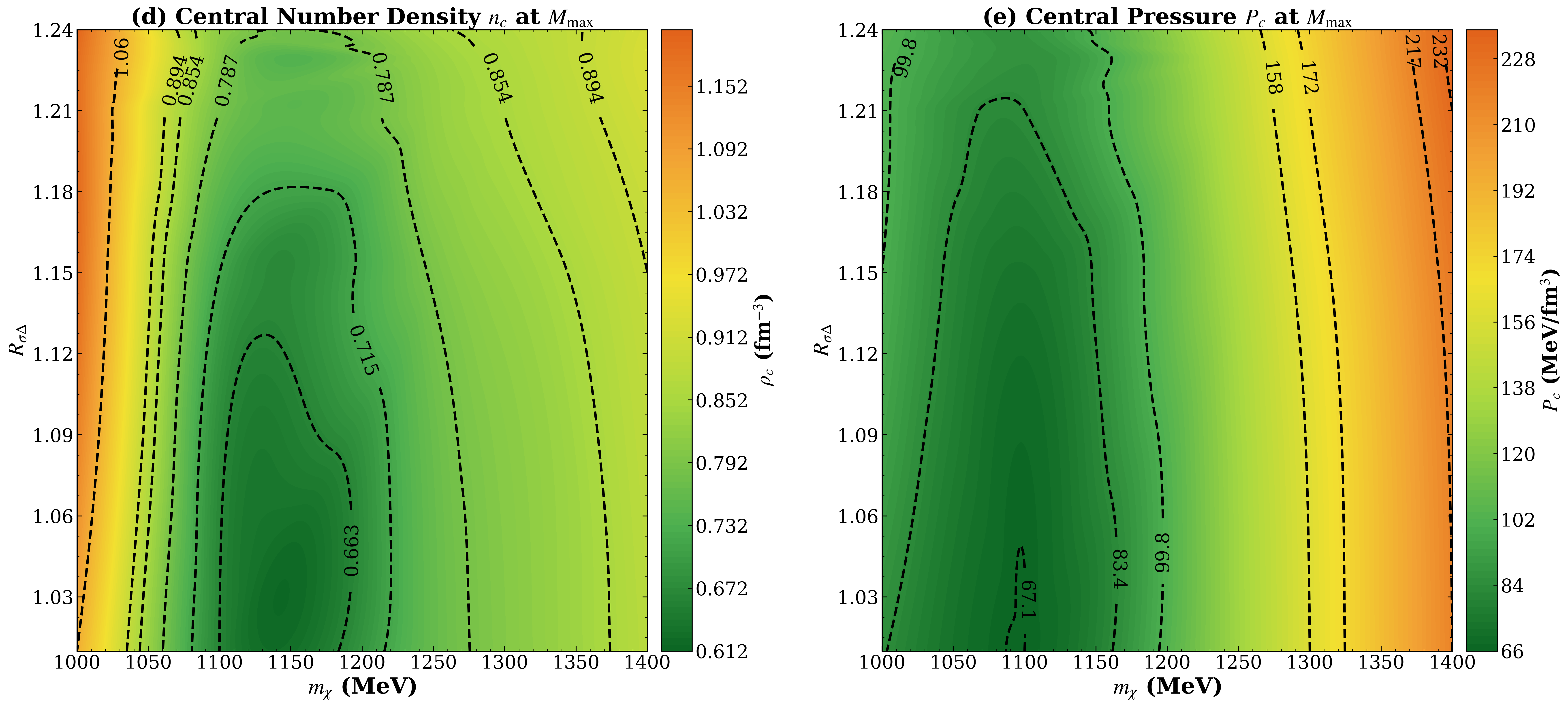}
    \captionof{figure}{NS observables in the $(m_\chi,R_{\sigma\Delta})$ parameter plane for fixed $R_{\omega\Delta}=1.1$, $R_{\rho\Delta}=1.0$ and $y_\chi=0.01$. The panels show (a) the canonical tidal deformability $\Lambda_{1.4}$, (b) the canonical radius $R_{1.4}$, (c) the maximum mass $M_{\max}$, (d) the central density $n_c^{\max}$ and (e) the central pressure $P_c^{\max}$ of the maximum-mass configuration. The contour structure demonstrates that the dark-baryon mass primarily controls the maximum mass, whereas the scalar $\Delta$ coupling introduces a secondary modification through the composition and stiffness of the high-density matter.}
    \label{fig:Contour3}
\end{figure*}

The combined dependence of the stellar observables on the dark-baryon mass $m_\chi$ and the scalar $\Delta$-meson coupling ratio $R_{\sigma\Delta}$ is summarized in Figure-\ref{fig:Contour3}. 
The calculations cover approximately
\begin{equation*}
1000\leq m_\chi\leq1400~{\rm MeV},
\qquad
1.00\leq R_{\sigma\Delta}\leq1.25,
\end{equation*}
while $R_{\omega\Delta}=1.1$, $R_{\rho\Delta}=1.0$ and
$y_\chi=0.01$ are held fixed. The five panels show the canonical tidal deformability, canonical radius, maximum mass as well as the central density and pressure of the maximum-mass configuration.
The resulting maps are insensitive to the Higgs mean-field correction at the numerical accuracy considered here. The variations displayed in Fig. \ref{fig:Contour3} are therefore controlled primarily by $m_\chi$ and $R_{\sigma\Delta}$.

The maximum-mass map in panel (c) exhibits an almost vertical contour structure, demonstrating that $M_{\max}$ is governed primarily by $m_\chi$ rather than by the variation of $R_{\sigma\Delta}$ over the considered interval. 
The maximum mass increases from approximately $0.8\,M_\odot$ near $m_\chi=1000$ MeV to about $2.0\,M_\odot$ at the upper end of the scanned mass range. 
This strong dependence follows from the density at which the dark state becomes energetically accessible. 
A light $\chi$ is populated comparatively early, converts a large fraction of the neutron component and produces severe softening of the EOS.
Increasing $m_\chi$ postpones the conversion to progressively higher density, thereby reducing the dark fraction and restoring part of the high-density pressure support.
Nevertheless, increasing $R_{\sigma\Delta}$ strengthens
the attractive scalar interaction in the $\Delta$ sector and produces a modest additional reduction of the high-density pressure. 
The region capable of supporting a $2~M_\odot$ NS is consequently restricted to a narrow band near
$m_\chi\simeq1.38$--$1.40$ GeV for the adopted values of the remaining couplings.
This viable interval refers specifically to the chemical-equilibrium limit considered here; incomplete $(n\leftrightarrow\chi)$ equilibration or additional dark-sector interactions could modify the resulting mass constraint.

\vt{The canonical radius shown in panel (b) varies approximately between $10.0$ and $12.9$ km. 
A rapid change occurs at the lower end of the
$m_\chi$ interval, where the dark baryon may already be populated in a $1.4\,M_\odot$ star. Once $m_\chi$ becomes sufficiently large that
the canonical configuration remains below the conversion threshold, the radius becomes considerably less sensitive to the dark-baryon mass and is controlled mainly by the hadronic composition and the $\Delta$-meson couplings.}

A similar threshold-driven behaviour is visible in the tidal deformability map of panel (a). 
The displayed values span a broad
interval, approximately $175\lesssim\Lambda_{1.4}\lesssim2425$.
The strongest variation is concentrated in the low-$m_\chi$ region, where the presence of $\chi$ changes the intermediate-density equation
of state sampled by a canonical star. At larger $m_\chi$, the contours become substantially flatter because the dark baryon appears only in
stars with central densities higher than that of the
$1.4\,M_\odot$ configuration. The canonical tidal response is then nearly insensitive to the dark sector even though the maximum-mass configuration may contain a dark core.

\begin{table*}
\centering
\caption{Summary of the neutron-star properties for the representative
matter compositions and dark-baryon masses considered in this work.
$Y$ denotes the hyperons. The complete calculations employ the
benchmark $y_\chi=0.01$; the corresponding Higgs-induced corrections
are negligible at the quoted numerical precision, as discussed in
Appendix A.}
\label{tab:summary}
\resizebox{0.75\textwidth}{!}{%
\begin{tabular}{lccccccc}
\hline
\hline
Composition & $m_\chi$ (MeV) & $M_{\max}$ ($M_\odot$) & $R_{\max}$ (km) &
$n_c^{\max}/n_0$ & $X_{\chi,c}^{\max}$ & \textcolor{black}{$R_{1.8}$ (km)} & \textcolor{black}{$n_c^{1.8}/n_0$}\\
\hline
$N$                              & --     & 2.48   & 12.0432     & 5.3726   & -- & \textcolor{black}{13.2439} & \textcolor{black}{2.6750} \\
$N+\chi$                         & 1250  & 2.00   & 13.1293   & 4.5014     & 0.3215 & \textcolor{black}{13.2439} & \textcolor{black}{2.6750} \\
$N+\chi$                         & 1300  & 2.11   & 13.0941   & 4.6547     & 0.2955 & \textcolor{black}{13.2439} & \textcolor{black}{2.6750} \\
$N+\chi$                         & 1400  & 2.27   & 12.9576   & 4.9547     & 0.2754 & \textcolor{black}{13.2439} & \textcolor{black}{2.6750} \\
$N+Y+\Delta+\chi$  & 1250  & 1.80   & 12.1784    & 5.3642   & 0.3038  & \textcolor{black}{12.1784} & \textcolor{black}{5.3642}  \\
$N+Y+\Delta+\chi$                & 1300  & 1.90   & 12.1064   & 5.5757    & 0.2786  & \textcolor{black}{12.2733} & \textcolor{black}{3.5983}  \\
$N+Y+\Delta+\chi$                & 1400  & 2.02   & 11.9079   & 5.9977    & 0.2323 & \textcolor{black}{12.2733} & \textcolor{black}{3.5983} \\
\hline
\hline
\end{tabular}}
\end{table*}

\vt{Panels (d) and (e) show the central density and pressure associated with the maximum-mass sequence. 
The central density varies over approximately
$0.61~{\rm fm}^{-3}\lesssim n_c^{\max} \lesssim 1.15~{\rm fm}^{-3}$,
while the corresponding central pressure spans approximately
$66~{\rm MeV\,fm^{-3}} \lesssim P_c^{\max} \lesssim 228~{\rm MeV\,fm^{-3}}$.
The large central densities obtained for smaller $m_\chi$ reflect the strong compressibility of the dark-softened EOS.
However, these configurations terminate at relatively low gravitational masses. 
As $m_\chi$ increases, the EOS recovers sufficient stiffness to support heavier stars, while the central pressure of the maximum-mass configuration rises substantially.
The nonmonotonic structures visible in the central-density and central-pressure contours arise from the competition among several thresholds. 
Changes in $m_\chi$ determine the onset of the dark baryon, whereas variations in $R_{\sigma\Delta}$ alter the appearance and abundance of the $\Delta$ resonances. 
The resulting rearrangement of the particle composition modifies the mesonic mean fields and, consequently, the central properties of the limiting stable configuration.}

\section{Conclusions} \label{sec:conclusions}

In this work, we have investigated the density-induced appearance of a neutral dark baryon $\chi$ in cold, charge-neutral and $\beta$-equilibrated NS matter containing hyperons and the complete $\Delta(1232)$ quartet.
The hadronic sector was described within the density-dependent DDME2 CDF model \cite{2005PhRvC..71b4312L}, while a scalar Higgs portal was additionally included as a possible interaction channel between the dark and hadronic sectors. A direct numerical evaluation shows that the corresponding mean-field correction is negligible for the EOS and stellar observables. The macroscopic effects reported here are thus driven predominantly by the density-induced $n\leftrightarrow\chi$ conversion and the associated rearrangement of the equilibrium composition.
In contrast to fixed-admixture approaches, the abundance of $\chi$ was obtained self-consistently from the chemical-equilibrium condition $\mu_n=\mu_\chi$, together with total baryon-number conservation and the coupled mean-field equations.
We stress that the $\chi$ masses considered here exceed the neutron mass; the present mechanism therefore describes the in-medium production of a heavier dark-sector baryon rather than a vacuum
neutron-decay solution to the neutron-lifetime anomaly.

\vt{Our results show that the appearance of conventional non-nucleonic degrees of freedom has a direct impact on the dark-baryon threshold (refer to Fig.\ref{fig:fraction}).
Hyperons and $\Delta-$resonances redistribute baryon number and modify the neutron chemical potential, thereby delaying the onset of $\chi$ and reducing its abundance relative to nucleonic+$\chi$ matter. 
This competition is also reflected in the radial composition of the stars. 
For the representative value $m_\chi=1400$ MeV, the dark component is absent from canonical-mass configurations and becomes appreciable only in sufficiently massive stars, where it forms a compact inner core surrounded by a broader hyperonic$-\Delta$ region.}

\vt{The macroscopic stellar properties are found to depend sensitively on $m_\chi$.
For nucleonic+$\chi$ matter, the maximum masses are $2.0011$, $2.1136$ and $2.2721\,M_\odot$ for $m_\chi=1250$, $1300$ and $1400$ MeV, respectively. 
After including hyperons and $\Delta-$resonances, the corresponding values decrease to $1.80$, $1.90$, and $2.02\,M_\odot$. 
Thus, within the adopted interaction scheme, the $m_\chi=1250$ and $1300$ MeV
benchmarks are incompatible with the conventional $2~M_\odot$ requirement, whereas the $m_\chi=1400$ MeV case only marginally satisfies it.
Within the assumed chemical-equilibrium limit and the adopted interaction scheme, the present results favor a comparatively heavy density-induced dark baryon, with the $2~M_\odot$ crossing occurring near $m_\chi\simeq1.38$ GeV for $R_{\sigma\Delta}=1.2$.
This estimate should, however, be regarded as model dependent rather than as a statistically derived lower bound.}


\vt{The canonical radius and tidal response display a considerably weaker dependence on $m_\chi$ in the heavy-dark-baryon regime. 
This occurs because the central densities of $\sim1.4\,M_\odot$ stars remain below the conversion threshold, so that $\chi$ is absent from these configurations. 
Separately, the calculated GW170817-like tidal trajectories
show greater sensitivity to the $\Delta$-meson coupling
than to the three sampled dark-baryon masses (in line with previous works-\cite{Thapa_2021, Thapa_2021_Bary, Parmar_2025}). 
By contrast, the maximum-mass configurations probe densities well above the dark threshold and therefore provide the strongest astrophysical sensitivity to the dark sector.}


\vt{The sound speed and adiabatic index further demonstrate the local softening associated with the appearance of new degrees of freedom.
The onset of $\chi$ produces a pronounced reduction in both $c_s^2$ and $\Gamma$, while additional structures arise from the successive population of hyperons and $\Delta$ resonances. 
These features provide a microscopic connection between the rearrangement of the particle composition and the reduction of the high-density pressure support.}


\textcolor{black}{Our dark-baryon mass constraints assume chemical equilibration during stellar evolution with negligible mixing-induced EOS corrections. Slower conversion could alter the composition and the inferred constraints.
The interaction model includes a scalar Higgs portal as a benchmark but omits additional vector-mediated interactions. Recent studies have explored vector portals between dark and hadronic matter \cite{issifu2026selfconsistenthiggsportalframeworkdark, 2026PhRvD.114d3027K} and repulsive interactions within the dark sector \cite{2026PhRvD.113f3014G}, demonstrating that these channels can modify chemical potentials, pressure and stellar structure. Although those frameworks differ from the present conversion-equilibrium model, they motivate examining how vector interactions would modify the equilibrium $\chi$ abundance and the resulting EOS. Consequently, the preference for heavier dark baryons obtained here is conditional on the adopted interaction model.
A microscopic calculation of the in-medium \(n\leftrightarrow\chi\) conversion rate, an extension including vector-mediated interactions, and a finite-temperature treatment are therefore important directions for future investigation.}
An especially interesting direction is the thermal evolution of these configurations, since the appearance of $\chi$ rearranges the neutron, proton, lepton, hyperon and $\Delta$ populations and may consequently modify the thresholds and emissivities of neutrino-emitting reactions, including nucleonic and hyperonic direct-Urca processes, as well as the specific heat and thermal conductivity of the stellar medium \cite{2022MNRAS.513.1820K, 2024Parti...7..179G, 2025PhRvD.112l3035Z}. 
Incorporating the present density-induced conversion mechanism into neutron-star cooling simulations could thus provide an independent probe of the dark-baryon onset beyond the mass--radius and tidal observables considered here. 
\textcolor{black}{A complementary probe of neutron conversion is provided by the thermal emission of old neutron stars. McKeen et al.~\cite{McKeen_2021_Mirror} showed that collision-induced neutron--mirror-neutron conversion can generate internal heating as nucleons refill the vacancies
created in their Fermi seas, allowing observations of cold NSs to constrain the mixing strength. This mechanism motivates ultraviolet, optical, and infrared
observations as probes of conversion-induced heating. Its application to the heavier dark baryons considered here would require a dedicated calculation of the in-medium conversion rate and associated energy release.}
In addition, dark-sector interactions may affect the late-time thermal balance of old NSs through kinetic-energy deposition or other heating channels, as discussed in captured-DM scenarios \cite{2017PhRvL.119m1801B, 2018PhRvD..97d3006R, 2018JCAP...09..018B, PhysRevD.77.023006}; in the present baryon-number-carrying conversion scenario, it would be particularly relevant to investigate whether out-of-equilibrium $n\leftrightarrow\chi$ reactions can generate or absorb heat and thereby leave observable signatures in the long-term surface-temperature evolution.

\begin{acknowledgements}
The authors thank the anonymous referee for the valuable suggestions provided, which have greatly contributed to enhancing the quality of this manuscript. NPK and VBT acknowledge the financial support from the Indian Space Research Organization (ISRO), Department of Space, Government of India through Project No. RAC-S/GU/2024/4/22.
They are also thankful to Athira S and Monika Sinha for vital and fruitful discussions.
\end{acknowledgements}

\appendix

\section*{APPENDIX A: Higgs-portal sector and numerical estimate} \label{appendix}

For completeness, we consider a scalar Higgs portal as a possible interaction channel between the dark baryon and the hadronic sector.
\textcolor{black}{We retain the scalar Higgs portal to quantify its correction to the conversion-equilibrium EOS. Additional vector-mediated interactions, explored in recent studies \cite{2026PhRvD.114d3027K, 2026PhRvD.113f3014G, issifu2026selfconsistenthiggsportalframeworkdark}, are outside the present model; their possible implications are discussed in Sec. \ref{sec:conclusions}.}
The corresponding Lagrangian density is given by \cite{2017PhRvD..96h3004P, 2019PhRvD..99d3016D, 2020MNRAS.495.4893D},
\begin{eqnarray}
\begin{aligned}
\mathcal{L}_{\rm Higgs}
=&
\frac{1}{2}\partial_\mu h\,\partial^\mu h
-\frac{1}{2}M_h^2h^2
+y_\chi h\bar{\chi}\chi
+\sum_{B}g_{hB}h\bar{\psi}_B\psi_B \\
& +\sum_{\Delta}g_{h\Delta}h\bar{\psi}_{\Delta \nu}\psi^\nu_\Delta,
\label{higgs}
\end{aligned}
\end{eqnarray}
where,
$g_{hj}={f_jm_j}/{v}$.
Here, \(M_h=125\) GeV is the mass of a Higgs particle. 
Previous Higgs-portal studies of dark-matter-admixed NSs have predominantly focused on nucleonic matter. Here we extend the framework to include the baryon octet and the complete $\Delta$ quartet.
For the adopted couplings, $\Sigma$ hyperons are absent over the density range considered.
Only the \(\Lambda\), \(\Xi^-\), \(\Xi^0\) and \(\Delta\)-resonances are energetically favourable, as discussed in \cite{Thapa_2021}. We have considered the nucleon-Higgs form factor as \(f=0.35\) \cite{2017PhRvD..96h3004P,2019PhRvD..99d3016D,2020MNRAS.495.4893D} and the same is taken for the \(\Delta\)-Higgs, baryon octet-Higgs interactions following the work \cite{Das_2021}, as their values are not known. 
\vt{The assumption $f_j=f$ should be regarded as a simplifying model input
rather than a fundamental relation.
We use this common form factor as a benchmark; the
Higgs-sector correction reported below refers to this
choice.}
\(v=246\) GeV is the vacuum expectation value of the Higgs field \cite{2019PhRvD..99d3016D, 2020MNRAS.495.4893D}. 
\textcolor{black}{The diagonal Higgs couplings mediate an additional scalar interaction between the hadronic and dark sectors. They do not generate the off-diagonal conversion operator in Eq.\eqref{conversion}. The $\chi$ abundance is determined by the separately imposed conversion-equilibrium condition.}

\vt{In the static mean-field approximation, the Higgs field satisfies
\begin{equation}
M_h^2 h
=
y_\chi n_\chi^s
+
\sum_{j= B,\Delta}g_{hj}n_j^s ,
\end{equation}
where,
\begin{equation}
n_\chi^s
=
\frac{1}{\pi^2}
\int_0^{k_{F\chi}}
\frac{m_\chi^* k^2\,dk}
{\sqrt{k^2+m_\chi^{*2}}}.
\end{equation}
}

For $m_\chi<M_h/2$, the Higgs portal permits the invisible decay
$h\rightarrow\chi\bar{\chi}$. Its vacuum partial width is
\begin{equation}
\Gamma(h\rightarrow\chi\bar{\chi})
=
\frac{y_\chi^2M_h}{8\pi}
\left(
1-\frac{4m_\chi^2}{M_h^2}
\right)^{3/2}.
\end{equation}
The corresponding invisible branching fraction is
\begin{equation}
{\rm Br}_{\rm inv}
=
\frac{\Gamma(h\rightarrow\chi\bar{\chi})}
{\Gamma_h^{\rm SM}+\Gamma(h\rightarrow\chi\bar{\chi})}.
\end{equation}
Using the experimental upper limit
${\rm Br}_{\rm inv}<0.107$ \cite{Aad_2023_1} and a Standard Model Higgs width of
approximately $4.1$ MeV \cite{Aad_2023_2} yields $y_\chi\lesssim 10^{-2}$ for the
dark-baryon masses considered here. 
We use $y_\chi=0.01$ as a rounded benchmark near this
limit to estimate the magnitude of the Higgs correction.

The Higgs field modifies the effective masses according to
\begin{equation}
m_j^*
=
m_j-g_{\sigma j}\sigma-g_{hj}h,
\qquad j=B,\Delta,
\end{equation}
and
\begin{equation}
m_\chi^*=m_\chi-y_\chi h.
\end{equation}
Its direct contributions to the pressure and energy density are
$-M_h^2h^2/2$ and $+M_h^2h^2/2$, respectively.

We solve the Higgs-field equation self-consistently together with the
hadronic mean-field and chemical-equilibrium equations. 
For $y_\chi=0.01$ and the adopted hadronic form factor, the resulting
Higgs-induced corrections to the effective-mass scale are negligible throughout the density range relevant to the stable stellar configurations. 
This scale is negligible compared with the strong-interaction mean fields and the characteristic fermionic energies, which are of order tens to hundreds of MeV. 
For the benchmark coupling $y_{\chi}=0.01$, the Higgs-induced correction to the dark-baryon effective mass is found to be of order $|y_{\chi}h| \sim 10^{-9}\ {\rm MeV}$ throughout the density range relevant to stable stellar configurations. 
This corresponds to a fractional correction $|\Delta m_{\chi}^{*}|/m_{\chi} \sim 10^{-12}$ for the dark-baryon masses considered here, and is therefore entirely negligible compared with the characteristic strong-interaction and fermionic energy scales.
Consequently, including the Higgs mean field produces no discernible change in the EOS, threshold densities, mass--radius sequences, or tidal observables at the numerical accuracy of the present calculation.
The principal dark-sector effect discussed in the main text therefore arises from the equilibrium population of $\chi$, rather than from Higgs-mediated interactions.

\bibliographystyle{apsrev4-1}

\bibliography{GW_ref,DM_ref,Kaons,Kaons2}

@ARTICLE{2021EPJC...81.1036B,
       author = {{Berezhiani}, Zurab and {Biondi}, Riccardo and {Mannarelli}, Massimo and {Tonelli}, Francesco},
        title = "{Neutron-mirror neutron mixing and neutron stars}",
      journal = {European Physical Journal C},
         year = 2021,
        month = nov,
       volume = {81},
       number = {11},
          eid = {1036},
        pages = {1036},
          doi = {10.1140/epjc/s10052-021-09806-1},
archivePrefix = {arXiv},
       eprint = {2012.15233},
 primaryClass = {astro-ph.HE},
       adsurl = {https://ui.adsabs.harvard.edu/abs/2021EPJC...81.1036B}
}

@ARTICLE{2026PhRvD.113f3014G,
       author = {{Grippa}, Francesco and {Lambiase}, Gaetano and {Poddar}, Tanmay Kumar},
        title = "{Scalar and vector dark matter admixed neutron stars with linear and quadratic couplings}",
      journal = {\prd},
         year = 2026,
        month = mar,
       volume = {113},
       number = {6},
          eid = {063014},
        pages = {063014},
          doi = {10.1103/6xxw-j8kt},
archivePrefix = {arXiv},
       eprint = {2407.16386},
 primaryClass = {hep-ph},
       adsurl = {https://ui.adsabs.harvard.edu/abs/2026PhRvD.113f3014G}
}

@ARTICLE{2026PhRvD.114d3027K,
       author = {{Kumar}, Deepak and {Mohapatra}, Ranjita K. and {Mishra}, Hiranmaya and {Patra}, Sudhanwa},
        title = "{Neutron star with dark matter using vector portal}",
      journal = {\prd},
         year = 2026,
        month = aug,
       volume = {114},
       number = {4},
          eid = {043027},
        pages = {043027},
          doi = {10.1103/fsh5-b73x},
archivePrefix = {arXiv},
       eprint = {2604.04560},
 primaryClass = {hep-ph},
       adsurl = {https://ui.adsabs.harvard.edu/abs/2026PhRvD.114d3027K}
}

@ARTICLE{2026PhRvC.113e5807D,
       author = {{Divaris}, M. and {Moustakidis}, Ch. C.},
        title = "{Neutron dark decay in neutron stars: The role of the symmetry energy}",
      journal = {\prc},
         year = 2026,
        month = may,
       volume = {113},
       number = {5},
          eid = {055807},
        pages = {055807},
          doi = {10.1103/23rn-52bw},
archivePrefix = {arXiv},
       eprint = {2508.21754},
 primaryClass = {nucl-th},
       adsurl = {https://ui.adsabs.harvard.edu/abs/2026PhRvC.113e5807D}
}

@ARTICLE{2024PhRvD.110h3003B,
       author = {{Bastero-Gil}, Mar and {Huertas-Rold{\'a}n}, Teresa and {Santos}, Daniel},
        title = "{Neutron decay anomaly, neutron stars, and dark matter}",
      journal = {\prd},
         year = 2024,
        month = oct,
       volume = {110},
       number = {8},
          eid = {083003},
        pages = {083003},
          doi = {10.1103/PhysRevD.110.083003},
archivePrefix = {arXiv},
       eprint = {2403.08666},
 primaryClass = {astro-ph.CO},
       adsurl = {https://ui.adsabs.harvard.edu/abs/2024PhRvD.110h3003B}
}

@ARTICLE{2021PhRvD.103k5002M,
       author = {{McKeen}, David and {Pospelov}, Maxim and {Raj}, Nirmal},
        title = "{Cosmological and astrophysical probes of dark baryons}",
      journal = {\prd},
         year = 2021,
        month = jun,
       volume = {103},
       number = {11},
          eid = {115002},
        pages = {115002},
          doi = {10.1103/PhysRevD.103.115002},
archivePrefix = {arXiv},
       eprint = {2012.09865},
 primaryClass = {hep-ph},
       adsurl = {https://ui.adsabs.harvard.edu/abs/2021PhRvD.103k5002M}
}

@ARTICLE{2018JCAP...09..018B,
       author = {{Bell}, Nicole F. and {Busoni}, Giorgio and {Robles}, Sandra},
        title = "{Heating up neutron stars with inelastic dark matter}",
      journal = {\jcap},
         year = 2018,
        month = sep,
       volume = {2018},
       number = {9},
          eid = {018},
        pages = {018},
          doi = {10.1088/1475-7516/2018/09/018},
archivePrefix = {arXiv},
       eprint = {1807.02840},
 primaryClass = {hep-ph},
       adsurl = {https://ui.adsabs.harvard.edu/abs/2018JCAP...09..018B}
}

@ARTICLE{2018PhRvD..97d3006R,
       author = {{Raj}, Nirmal and {Tanedo}, Philip and {Yu}, Hai-Bo},
        title = "{Neutron stars at the dark matter direct detection frontier}",
      journal = {\prd},
         year = 2018,
        month = feb,
       volume = {97},
       number = {4},
          eid = {043006},
        pages = {043006},
          doi = {10.1103/PhysRevD.97.043006},
archivePrefix = {arXiv},
       eprint = {1707.09442},
 primaryClass = {hep-ph},
       adsurl = {https://ui.adsabs.harvard.edu/abs/2018PhRvD..97d3006R}
}

@ARTICLE{2017PhRvL.119m1801B,
       author = {{Baryakhtar}, Masha and {Bramante}, Joseph and {Li}, Shirley Weishi and {Linden}, Tim and {Raj}, Nirmal},
        title = "{Dark Kinetic Heating of Neutron Stars and an Infrared Window on WIMPs, SIMPs, and Pure Higgsinos}",
      journal = {\prl},
         year = 2017,
        month = sep,
       volume = {119},
       number = {13},
          eid = {131801},
        pages = {131801},
          doi = {10.1103/PhysRevLett.119.131801},
archivePrefix = {arXiv},
       eprint = {1704.01577},
 primaryClass = {hep-ph},
       adsurl = {https://ui.adsabs.harvard.edu/abs/2017PhRvL.119m1801B}
}

@ARTICLE{2024Parti...7..179G,
       author = {{Giangrandi}, Edoardo and {{\'A}vila}, Afonso and {Sagun}, Violetta and {Ivanytskyi}, Oleksii and {Provid{\^e}ncia}, Constan{\c{c}}a},
        title = "{The Impact of Asymmetric Dark Matter on the Thermal Evolution of Nucleonic and Hyperonic Compact Stars}",
      journal = {Particles},
         year = 2024,
        month = feb,
       volume = {7},
       number = {1},
        pages = {179-200},
          doi = {10.3390/particles7010010},
archivePrefix = {arXiv},
       eprint = {2401.03295},
 primaryClass = {astro-ph.HE},
       adsurl = {https://ui.adsabs.harvard.edu/abs/2024Parti...7..179G}
}

@ARTICLE{2022MNRAS.513.1820K,
       author = {{Kumar}, Ankit and {Das}, H.~C. and {Patra}, S.~K.},
        title = "{Thermal relaxation of dark matter admixed neutron star}",
      journal = {\mnras},
         year = 2022,
        month = jun,
       volume = {513},
       number = {2},
        pages = {1820-1833},
          doi = {10.1093/mnras/stac1013},
archivePrefix = {arXiv},
       eprint = {2203.02132},
 primaryClass = {astro-ph.HE},
       adsurl = {https://ui.adsabs.harvard.edu/abs/2022MNRAS.513.1820K}
}

@ARTICLE{2025PhRvD.112l3035Z,
       author = {{Zhou}, B.~X. and {Das}, H.~C. and {Wei}, J.~B. and {Burgio}, G.~F. and {Li}, Z.~H. and {Schulze}, H.-J.},
        title = "{Cooling of dark neutron stars}",
      journal = {\prd},
         year = 2025,
        month = dec,
       volume = {112},
       number = {12},
          eid = {123035},
        pages = {123035},
          doi = {10.1103/nzq6-llbf},
archivePrefix = {arXiv},
       eprint = {2508.09704},
 primaryClass = {astro-ph.HE},
       adsurl = {https://ui.adsabs.harvard.edu/abs/2025PhRvD.112l3035Z}
}

@article{3zy4-h2ty,
  title = {Properties of dark matter admixed neutron star within relativistic mean field model},
  author = {Iqbal, Tamanna and Kumaran, Yashmitha and Bhagwat, A. and Sharma, B. K.},
  journal = {Phys. Rev. D},
  volume = {113},
  issue = {2},
  pages = {023030},
  numpages = {9},
  year = {2026},
  month = {Jan},
  publisher = {American Physical Society},
  doi = {10.1103/3zy4-h2ty},
  url = {https://link.aps.org/doi/10.1103/3zy4-h2ty}
}

@ARTICLE{2024PhRvD.110f3001K,
       author = {{Kumar}, Ankit and {Sotani}, Hajime},
        title = "{Constraints on the parameter space in dark matter admixed neutron stars}",
      journal = {\prd},
         year = 2024,
        month = sep,
       volume = {110},
       number = {6},
          eid = {063001},
        pages = {063001},
          doi = {10.1103/PhysRevD.110.063001},
archivePrefix = {arXiv},
       eprint = {2408.15312},
 primaryClass = {astro-ph.HE},
       adsurl = {https://ui.adsabs.harvard.edu/abs/2024PhRvD.110f3001K}
}

@misc{kumar2026slowlyrotatingtwofluidneutron,
      title={Slowly Rotating Two-Fluid Neutron Stars: Coupled Frame-Dragging, Inertia Splitting, and Universal Relations}, 
      author={Ankit Kumar and Hajime Sotani},
      year={2026},
      eprint={2603.12613},
      archivePrefix={arXiv},
      primaryClass={astro-ph.HE},
      url={https://arxiv.org/abs/2603.12613}, 
}

@misc{das2020darkmatteradmixedneutron,
      title={Dark matter admixed neutron star properties in the light of gravitational wave observations: a two fluid approach}, 
      author={Arpan Das and Tuhin Malik and Alekha C. Nayak},
      year={2020},
      eprint={2011.01318},
      archivePrefix={arXiv},
      primaryClass={nucl-th},
      url={https://arxiv.org/abs/2011.01318}, 
}

@ARTICLE{2021PhRvD.103d3019G,
       author = {{Garani}, Raghuveer and {Gupta}, Aritra and {Raj}, Nirmal},
        title = "{Observing the thermalization of dark matter in neutron stars}",
      journal = {\prd},
         year = 2021,
        month = feb,
       volume = {103},
       number = {4},
          eid = {043019},
        pages = {043019},
          doi = {10.1103/PhysRevD.103.043019},
archivePrefix = {arXiv},
       eprint = {2009.10728},
 primaryClass = {hep-ph},
       adsurl = {https://ui.adsabs.harvard.edu/abs/2021PhRvD.103d3019G}
}

@ARTICLE{issifu2026darkmatterheatingevolving,
       author = {{Issifu}, Adamu and {Thakur}, Prashant and {Karkevandi}, Davood Rafiei and {da Silva}, Franciele M. and {Menezes}, D{\'e}bora P. and {Lim}, Y. and {Frederico}, Tobias},
        title = "{Dark matter heating in evolving protoneutron stars: A two-fluid approach}",
      journal = {\prd},
         year = 2026,
        month = may,
       volume = {113},
       number = {10},
          eid = {103042},
        pages = {103042},
          doi = {10.1103/bhm3-jzjq},
archivePrefix = {arXiv},
       eprint = {2511.07567},
 primaryClass = {astro-ph.HE},
       adsurl = {https://ui.adsabs.harvard.edu/abs/2026PhRvD.113j3042I}
}

@ARTICLE{Mariani_2023,
       author = {{Mariani}, Mauro and {Albertus}, Conrado and {Alessandroni}, M. del Rosario and {Orsaria}, Milva G. and {P{\'e}rez-Garc{\'\i}a}, M. {\'A}ngeles and {Ranea-Sandoval}, Ignacio F.},
        title = "{Constraining self-interacting fermionic dark matter in admixed neutron stars using multimessenger astronomy}",
      journal = {\mnras},
         year = 2024,
        month = jan,
       volume = {527},
       number = {3},
        pages = {6795-6806},
          doi = {10.1093/mnras/stad3658},
archivePrefix = {arXiv},
       eprint = {2311.14004},
 primaryClass = {astro-ph.HE},
       adsurl = {https://ui.adsabs.harvard.edu/abs/2024MNRAS.527.6795M}
}

@article{Issifu_2026,
   title={Self-consistent single-fluid framework for neutron stars admixed with mirror dark matter},
   volume={114},
   ISSN={2470-0029},
   url={http://dx.doi.org/10.1103/ftgk-cm86},
   DOI={10.1103/ftgk-cm86},
   number={6},
   journal={Physical Review D},
   publisher={American Physical Society (APS)},
   author={Issifu, Adamu and Providência, Constança and da Silva, Franciele M. and Menezes, Débora P. and Frederico, Tobias},
   year={2026},
   month=Sept }

@misc{issifu2026selfconsistenthiggsportalframeworkdark,
      title={A self-consistent Higgs-portal framework for dark matter--admixed neutron stars: Collider-motivated benchmarks meet multimessenger constraints}, 
      author={Adamu Issifu},
      year={2026},
      eprint={2607.14979},
      archivePrefix={arXiv},
      primaryClass={astro-ph.HE},
      url={https://arxiv.org/abs/2607.14979}, 
}

@article{Lopes_2023,
   title={Baryon coupling scheme in a unified SU(3) and SU(6) symmetry formalism},
   volume={107},
   ISSN={2470-0029},
   url={http://dx.doi.org/10.1103/PhysRevD.107.036011},
   DOI={10.1103/physrevd.107.036011},
   number={3},
   journal={Physical Review D},
   publisher={American Physical Society (APS)},
   author={Lopes, Luiz L. and Marquez, Kauan D. and Menezes, Débora P.},
   year={2023},
   month=Feb }

@article{LIGO_Virgo2017a,
  title={Multi-messenger observations of a binary neutron star merger},
  author={Abbott, B. P. and Abbott, R. and Adhikari, R. X. and Ananyeva, A. and Anderson, S. B. and others},
  journal={ApJL},
  volume={848},
  number={2},
  pages={L12},
  year={2017},
  publisher={American Astronomical Society}
}

@article{LIGO_Virgo2017b,
  title={Gravitational waves and gamma-rays from a binary neutron star merger: GW170817 and GRB 170817A},
  author={Abbott, B. P. and Abbott, R. and Abbott, T. D. and Acernese, F. and Ackley, K. and others},
  journal={ApJL},
  volume={848},
  number={2},
  pages={L13},
  year={2017},
  publisher={IOP Publishing}
}

@article{LIGO_Virgo2017c,
  title = {GW170817: Observation of Gravitational Waves from a Binary Neutron Star Inspiral},
  author = {Abbott, B. P. and Abbott, R. and Abbott, T. D. and Acernese, F. and Ackley, K. and others},
  collaboration = {LIGO Scientific Collaboration and Virgo Collaboration},
  journal = {PhRvL},
  volume = {119},
  issue = {16},
  pages = {161101},
  numpages = {18},
  year = {2017},
  month = {Oct},
  publisher = {American Physical Society}
}

@article{Romani_2021,
   title={PSR J1810+1744: Companion Darkening and a Precise High Neutron Star Mass},
   volume={908},
   ISSN={2041-8213},
   url={http://dx.doi.org/10.3847/2041-8213/abe2b4},
   DOI={10.3847/2041-8213/abe2b4},
   number={2},
   journal={The Astrophysical Journal Letters},
   publisher={American Astronomical Society},
   author={Romani, Roger W. and Kandel, D. and Filippenko, Alexei V. and Brink, Thomas G. and Zheng, WeiKang},
   year={2021},
   month=Feb, pages={L46} }

@article{Thapa_2021_Bary,
   title={Baryonic dense matter in view of gravitational-wave observations},
   volume={507},
   ISSN={1365-2966},
   url={http://dx.doi.org/10.1093/mnras/stab2327},
   DOI={10.1093/mnras/stab2327},
   number={2},
   journal={Monthly Notices of the Royal Astronomical Society},
   publisher={Oxford University Press (OUP)},
   author={Thapa, Vivek Baruah and Kumar, Anil and Sinha, Monika},
   year={2021},
   month=Aug, pages={2991–3004} }

@ARTICLE{2018PhRvL.121f1802M,
       author = {{McKeen}, David and {Nelson}, Ann E. and {Reddy}, Sanjay and {Zhou}, Dake},
        title = "{Neutron Stars Exclude Light Dark Baryons}",
      journal = {\prl},
         year = 2018,
        month = aug,
       volume = {121},
       number = {6},
          eid = {061802},
        pages = {061802},
          doi = {10.1103/PhysRevLett.121.061802},
archivePrefix = {arXiv},
       eprint = {1802.08244},
 primaryClass = {hep-ph},
       adsurl = {https://ui.adsabs.harvard.edu/abs/2018PhRvL.121f1802M}
}

@ARTICLE{2018PhRvL.121f1801B,
       author = {{Baym}, Gordon and {Beck}, D.~H. and {Geltenbort}, Peter and {Shelton}, Jessie},
        title = "{Testing Dark Decays of Baryons in Neutron Stars}",
      journal = {\prl},
         year = 2018,
        month = aug,
       volume = {121},
       number = {6},
          eid = {061801},
        pages = {061801},
          doi = {10.1103/PhysRevLett.121.061801},
archivePrefix = {arXiv},
       eprint = {1802.08282},
 primaryClass = {hep-ph},
       adsurl = {https://ui.adsabs.harvard.edu/abs/2018PhRvL.121f1801B}
}

@ARTICLE{2024PhR..1052....1B,
       author = {{Bramante}, Joseph and {Raj}, Nirmal},
        title = "{Dark matter in compact stars}",
      journal = {\physrep},
         year = 2024,
        month = feb,
       volume = {1052},
        pages = {1-48},
          doi = {10.1016/j.physrep.2023.12.001},
archivePrefix = {arXiv},
       eprint = {2307.14435},
 primaryClass = {hep-ph},
       adsurl = {https://ui.adsabs.harvard.edu/abs/2024PhR..1052....1B}
}

@ARTICLE{2020JCAP...09..028B,
       author = {{Bell}, Nicole F. and {Busoni}, Giorgio and {Robles}, Sandra and {Virgato}, Michael},
        title = "{Improved treatment of dark matter capture in neutron stars}",
      journal = {\jcap},
         year = 2020,
        month = sep,
       volume = {2020},
       number = {9},
          eid = {028},
        pages = {028},
          doi = {10.1088/1475-7516/2020/09/028},
archivePrefix = {arXiv},
       eprint = {2004.14888},
 primaryClass = {hep-ph},
       adsurl = {https://ui.adsabs.harvard.edu/abs/2020JCAP...09..028B}
}

@ARTICLE{2019JCAP...06..054B,
       author = {{Bell}, Nicole F. and {Busoni}, Giorgio and {Robles}, Sandra},
        title = "{Capture of leptophilic dark matter in neutron stars}",
      journal = {\jcap},
         year = 2019,
        month = jun,
       volume = {2019},
       number = {6},
          eid = {054},
        pages = {054},
          doi = {10.1088/1475-7516/2019/06/054},
archivePrefix = {arXiv},
       eprint = {1904.09803},
 primaryClass = {hep-ph},
       adsurl = {https://ui.adsabs.harvard.edu/abs/2019JCAP...06..054B}
}

@ARTICLE{2018PhRvD..97l3007E,
       author = {{Ellis}, John and {H{\"u}tsi}, Gert and {Kannike}, Kristjan and {Marzola}, Luca and {Raidal}, Martti and {Vaskonen}, Ville},
        title = "{Dark matter effects on neutron star properties}",
      journal = {\prd},
         year = 2018,
        month = jun,
       volume = {97},
       number = {12},
          eid = {123007},
        pages = {123007},
          doi = {10.1103/PhysRevD.97.123007},
archivePrefix = {arXiv},
       eprint = {1804.01418},
 primaryClass = {astro-ph.CO},
       adsurl = {https://ui.adsabs.harvard.edu/abs/2018PhRvD..97l3007E}
}

@ARTICLE{2020MNRAS.495.4893D,
       author = {{Das}, H.~C. and {Kumar}, Ankit and {Kumar}, Bharat and {Biswal}, S.~K. and {Nakatsukasa}, Takashi and {Li}, Ang and {Patra}, S.~K.},
        title = "{Effects of dark matter on the nuclear and neutron star matter}",
      journal = {\mnras},
         year = 2020,
        month = jul,
       volume = {495},
       number = {4},
        pages = {4893-4903},
          doi = {10.1093/mnras/staa1435},
archivePrefix = {arXiv},
       eprint = {2002.00594},
 primaryClass = {nucl-th},
       adsurl = {https://ui.adsabs.harvard.edu/abs/2020MNRAS.495.4893D}
}

@ARTICLE{2017PhRvD..96h3004P,
       author = {{Panotopoulos}, Grigorios and {Lopes}, Il{\'\i}dio},
        title = "{Dark matter effect on realistic equation of state in neutron stars}",
      journal = {\prd},
         year = 2017,
        month = oct,
       volume = {96},
       number = {8},
          eid = {083004},
        pages = {083004},
          doi = {10.1103/PhysRevD.96.083004},
archivePrefix = {arXiv},
       eprint = {1709.06312},
 primaryClass = {hep-ph},
       adsurl = {https://ui.adsabs.harvard.edu/abs/2017PhRvD..96h3004P}
}

@ARTICLE{2024JCAP...12..042T,
       author = {{Thakur}, Pratik and {Kumar}, Anil and {Thapa}, Vivek Baruah and {Parmar}, Vishal and {Sinha}, Monika},
        title = "{Exploring non-radial oscillation modes in dark matter admixed neutron stars}",
      journal = {\jcap},
         year = 2024,
        month = dec,
       volume = {2024},
       number = {12},
          eid = {042},
        pages = {042},
          doi = {10.1088/1475-7516/2024/12/042},
archivePrefix = {arXiv},
       eprint = {2406.07470},
 primaryClass = {astro-ph.HE},
       adsurl = {https://ui.adsabs.harvard.edu/abs/2024JCAP...12..042T}
}

@ARTICLE{2025MNRAS.544.3549S,
       author = {{Shirke}, Swarnim and {Chatterjee}, Debarati and {Jaikumar}, Prashanth},
        title = "{g-mode oscillations of dark matter admixed neutron stars}",
      journal = {\mnras},
         year = 2025,
        month = dec,
       volume = {544},
       number = {4},
        pages = {3549-3561},
          doi = {10.1093/mnras/staf1859},
archivePrefix = {arXiv},
       eprint = {2506.18892},
 primaryClass = {gr-qc},
       adsurl = {https://ui.adsabs.harvard.edu/abs/2025MNRAS.544.3549S}
}

@ARTICLE{2021MNRAS.507.4053D,
       author = {{Das}, H.~C. and {Kumar}, Ankit and {Patra}, S.~K.},
        title = "{Effects of dark matter on the in-spiral properties of the binary neutron stars}",
      journal = {\mnras},
         year = 2021,
        month = nov,
       volume = {507},
       number = {3},
        pages = {4053-4060},
          doi = {10.1093/mnras/stab2387},
archivePrefix = {arXiv},
       eprint = {2104.01815},
 primaryClass = {astro-ph.HE},
       adsurl = {https://ui.adsabs.harvard.edu/abs/2021MNRAS.507.4053D}
}

@ARTICLE{2019PhRvD..99d3016D,
       author = {{Das}, Arpan and {Malik}, Tuhin and {Nayak}, Alekha C.},
        title = "{Confronting nuclear equation of state in the presence of dark matter using GW170817 observation in relativistic mean field theory approach}",
      journal = {\prd},
         year = 2019,
        month = feb,
       volume = {99},
       number = {4},
          eid = {043016},
        pages = {043016},
          doi = {10.1103/PhysRevD.99.043016},
archivePrefix = {arXiv},
       eprint = {1807.10013},
 primaryClass = {hep-ph},
       adsurl = {https://ui.adsabs.harvard.edu/abs/2019PhRvD..99d3016D}
}

@ARTICLE{2011PhRvD..84j7301L,
       author = {{Leung}, S.-C. and {Chu}, M.-C. and {Lin}, L.-M.},
        title = "{Dark-matter admixed neutron stars}",
      journal = {\prd},
         year = 2011,
        month = nov,
       volume = {84},
       number = {10},
          eid = {107301},
        pages = {107301},
          doi = {10.1103/PhysRevD.84.107301},
archivePrefix = {arXiv},
       eprint = {1111.1787},
 primaryClass = {astro-ph.CO},
       adsurl = {https://ui.adsabs.harvard.edu/abs/2011PhRvD..84j7301L}
}

@ARTICLE{1989PhRvD..40.3221G,
       author = {{Goldman}, Itzhak and {Nussinov}, Shmuel},
        title = "{Weakly interacting massive particles and neutron stars}",
      journal = {\prd},
         year = 1989,
        month = nov,
       volume = {40},
       number = {10},
        pages = {3221-3230},
          doi = {10.1103/PhysRevD.40.3221},
       adsurl = {https://ui.adsabs.harvard.edu/abs/1989PhRvD..40.3221G}
}

@article{PhysRevD.77.023006,
  title = {WIMP annihilation and cooling of neutron stars},
  author = {Kouvaris, Chris},
  journal = {Phys. Rev. D},
  volume = {77},
  issue = {2},
  pages = {023006},
  numpages = {9},
  year = {2008},
  month = {Jan},
  publisher = {American Physical Society},
  doi = {10.1103/PhysRevD.77.023006},
  url = {https://link.aps.org/doi/10.1103/PhysRevD.77.023006}
}

@ARTICLE{2008PhRvD..77d3515B,
       author = {{Bertone}, Gianfranco and {Fairbairn}, Malcolm},
        title = "{Compact stars as dark matter probes}",
      journal = {\prd},
         year = 2008,
        month = feb,
       volume = {77},
       number = {4},
          eid = {043515},
        pages = {043515},
          doi = {10.1103/PhysRevD.77.043515},
archivePrefix = {arXiv},
       eprint = {0709.1485},
 primaryClass = {astro-ph},
       adsurl = {https://ui.adsabs.harvard.edu/abs/2008PhRvD..77d3515B}
}

@article{Fornal_2018,
   title={Dark Matter Interpretation of the Neutron Decay Anomaly},
   volume={120},
   ISSN={1079-7114},
   url={http://dx.doi.org/10.1103/PhysRevLett.120.191801},
   DOI={10.1103/physrevlett.120.191801},
   number={19},
   journal={Physical Review Letters},
   publisher={American Physical Society (APS)},
   author={Fornal, Bartosz and Grinstein, Benjamín},
   year={2018},
   month=May }

@article{Das_2021,
   title={Impacts of dark matter on the 
<mml:math xmlns:mml=“http://www.w3.org/1998/Math/MathML” display=“inline”><mml:mi>f</mml:mi></mml:math>
-mode oscillation of hyperon star},
   volume={104},
   ISSN={2470-0029},
   url={http://dx.doi.org/10.1103/PhysRevD.104.123006},
   DOI={10.1103/physrevd.104.123006},
   number={12},
   journal={Physical Review D},
   publisher={American Physical Society (APS)},
   author={Das, H.C. and Kumar, Ankit and Biswal, S. K. and Patra, S. K.},
   year={2021},
   month=Dec }

@ARTICLE{2016PhRvC..94c5804F,
       author = {{Fortin}, M. and {Provid{\^e}ncia}, C. and {Raduta}, Ad. R. and {Gulminelli}, F. and {Zdunik}, J.~L. and {Haensel}, P. and {Bejger}, M.},
        title = "{Neutron star radii and crusts: Uncertainties and unified equations of state}",
      journal = {\prc},
         year = 2016,
        month = sep,
       volume = {94},
       number = {3},
          eid = {035804},
        pages = {035804},
          doi = {10.1103/PhysRevC.94.035804},
archivePrefix = {arXiv},
       eprint = {1604.01944},
 primaryClass = {astro-ph.SR},
       adsurl = {https://ui.adsabs.harvard.edu/abs/2016PhRvC..94c5804F}
}

@ARTICLE{Lattimer2016PhR,
       author = {{Lattimer}, James M. and {Prakash}, Madappa},
        title = "{The equation of state of hot, dense matter and neutron stars}",
      journal = {\physrep},
         year = 2016,
        month = mar,
       volume = {621},
        pages = {127-164},
          doi = {10.1016/j.physrep.2015.12.005},
archivePrefix = {arXiv},
       eprint = {1512.07820},
 primaryClass = {astro-ph.SR},
       adsurl = {https://ui.adsabs.harvard.edu/abs/2016PhR...621..127L}
}

@BOOK{1996cost.book.....G,
       author = {{Glendenning}, Norman K.},
        title = "{Compact Stars}",
         year = 1996,
       adsurl = {https://ui.adsabs.harvard.edu/abs/1996cost.book.....G}
}

@ARTICLE{2020NatAs...4...72C,
       author = {{Cromartie}, H.~T. and {Fonseca}, E. and {Ransom}, S.~M. and
         {Demorest}, P.~B. and {Arzoumanian}, Z. and {Blumer}, H. and
         {Brook}, P.~R. and {DeCesar}, M.~E. and {Dolch}, T. and {Ellis}, J.~A. and
         {Ferdman}, R.~D. and {Ferrara}, E.~C. and {Garver-Daniels}, N. and
         {Gentile}, P.~A. and {Jones}, M.~L. and {Lam}, M.~T. and
         {Lorimer}, D.~R. and {Lynch}, R.~S. and {McLaughlin}, M.~A. and
         {Ng}, C. and {Nice}, D.~J. and {Pennucci}, T.~T. and {Spiewak}, R. and
         {Stairs}, I.~H. and {Stovall}, K. and {Swiggum}, J.~K. and {Zhu}, W.~W.},
        title = "{Relativistic Shapiro delay measurements of an extremely massive millisecond pulsar}",
      journal = {Nature Astronomy},
         year = 2020,
        month = jan,
       volume = {4},
        pages = {72-76},
          doi = {10.1038/s41550-019-0880-2},
archivePrefix = {arXiv},
       eprint = {1904.06759},
 primaryClass = {astro-ph.HE},
       adsurl = {https://ui.adsabs.harvard.edu/abs/2020NatAs...4...72C}
}

@ARTICLE{2020ApJ...896L..44A,
       author = {{Abbott}, R. and {Abbott}, T.~D. and {Abraham}, S. and {Acernese}, F. and
         {Ackley}, K. and {Adams}, C. and {Adhikari}, R.~X. and {Adya}, V.~B. and
         {Affeldt}, C. and {Agathos}, M. and et al.},
        title = "{GW190814: Gravitational Waves from the Coalescence of a 23 Solar Mass Black Hole with a 2.6 Solar Mass Compact Object}",
      journal = {\apjl},
         year = 2020,
        month = jun,
       volume = {896},
       number = {2},
          eid = {L44},
        pages = {L44},
          doi = {10.3847/2041-8213/ab960f},
archivePrefix = {arXiv},
       eprint = {2006.12611},
 primaryClass = {astro-ph.HE},
       adsurl = {https://ui.adsabs.harvard.edu/abs/2020ApJ...896L..44A}
}

@ARTICLE{1995PhRvC..52.3470K,
       author = {{Knorren}, R. and {Prakash}, M. and {Ellis}, P.~J.},
        title = "{Strangeness in hadronic stellar matter}",
      journal = {\prc},
         year = 1995,
        month = dec,
       volume = {52},
       number = {6},
        pages = {3470-3482},
          doi = {10.1103/PhysRevC.52.3470},
archivePrefix = {arXiv},
       eprint = {nucl-th/9506016},
 primaryClass = {nucl-th},
       adsurl = {https://ui.adsabs.harvard.edu/abs/1995PhRvC..52.3470K}
}

@ARTICLE{2019ApJ...887L..24M,
       author = {{Miller}, M.~C. and {Lamb}, F.~K. and {Dittmann}, A.~J. and
         {Bogdanov}, S. and {Arzoumanian}, Z. and {Gendreau}, K.~C. and
         {Guillot}, S. and {Harding}, A.~K. and {Ho}, W.~C.~G. and
         {Lattimer}, J.~M. and {Ludlam}, R.~M. and {Mahmoodifar}, S. and
         {Morsink}, S.~M. and {Ray}, P.~S. and {Strohmayer}, T.~E. and
         {Wood}, K.~S. and {Enoto}, T. and {Foster}, R. and {Okajima}, T. and
         {Prigozhin}, G. and {Soong}, Y.},
        title = "{PSR J0030+0451 Mass and Radius from NICER Data and Implications for the Properties of Neutron Star Matter}",
      journal = {\apjl},
         year = 2019,
        month = dec,
       volume = {887},
       number = {1},
          eid = {L24},
        pages = {L24},
          doi = {10.3847/2041-8213/ab50c5},
archivePrefix = {arXiv},
       eprint = {1912.05705},
 primaryClass = {astro-ph.HE},
       adsurl = {https://ui.adsabs.harvard.edu/abs/2019ApJ...887L..24M}
}

@ARTICLE{2019ApJ...887L..21R,
       author = {{Riley}, T.~E. and {Watts}, A.~L. and {Bogdanov}, S. and {Ray}, P.~S. and
         {Ludlam}, R.~M. and {Guillot}, S. and {Arzoumanian}, Z. and
         {Baker}, C.~L. and {Bilous}, A.~V. and {Chakrabarty}, D. and
         {Gendreau}, K.~C. and {Harding}, A.~K. and {Ho}, W.~C.~G. and
         {Lattimer}, J.~M. and {Morsink}, S.~M. and {Strohmayer}, T.~E.},
        title = "{A NICER View of PSR J0030+0451: Millisecond Pulsar Parameter Estimation}",
      journal = {\apjl},
         year = 2019,
        month = dec,
       volume = {887},
       number = {1},
          eid = {L21},
        pages = {L21},
          doi = {10.3847/2041-8213/ab481c},
archivePrefix = {arXiv},
       eprint = {1912.05702},
 primaryClass = {astro-ph.HE},
       adsurl = {https://ui.adsabs.harvard.edu/abs/2019ApJ...887L..21R}
}

@ARTICLE{2007PrPNP..59...94W,
       author = {{Weber}, F. and {Negreiros}, R. and {Rosenfield}, P. and {Stejner}, M.},
        title = "{Pulsars as astrophysical laboratories for nuclear and particle physics}",
      journal = {Progress in Particle and Nuclear Physics},
         year = 2007,
        month = jul,
       volume = {59},
       number = {1},
        pages = {94-113},
          doi = {10.1016/j.ppnp.2006.12.008},
archivePrefix = {arXiv},
       eprint = {astro-ph/0612054},
 primaryClass = {astro-ph},
       adsurl = {https://ui.adsabs.harvard.edu/abs/2007PrPNP..59...94W}
}

@ARTICLE{1985ApJ...293..470G,
       author = {{Glendenning}, N.~K.},
        title = "{Neutron stars are giant hypernuclei ?}",
      journal = {\apj},
         year = 1985,
        month = jun,
       volume = {293},
        pages = {470-493},
          doi = {10.1086/163253},
       adsurl = {https://ui.adsabs.harvard.edu/abs/1985ApJ...293..470G}
}

@ARTICLE{2005PhRvC..71b4312L,
       author = {{Lalazissis}, G.~A. and {Nik{\v{s}}i{\'c}}, T. and {Vretenar}, D. and
         {Ring}, P.},
        title = "{New relativistic mean-field interaction with density-dependent meson-nucleon couplings}",
      journal = {\prc},
         year = 2005,
        month = feb,
       volume = {71},
       number = {2},
          eid = {024312},
        pages = {024312},
          doi = {10.1103/PhysRevC.71.024312},
       adsurl = {https://ui.adsabs.harvard.edu/abs/2005PhRvC..71b4312L}
}

@ARTICLE{1971ApJ...170..299B,
       author = {{Baym}, Gordon and {Pethick}, Christopher and {Sutherland}, Peter},
        title = "{The Ground State of Matter at High Densities: Equation of State and Stellar Models}",
      journal = {\apj},
         year = 1971,
        month = dec,
       volume = {170},
        pages = {299},
          doi = {10.1086/151216},
       adsurl = {https://ui.adsabs.harvard.edu/abs/1971ApJ...170..299B}
}

@ARTICLE{2010Natur.467.1081D,
       author = {{Demorest}, P.~B. and {Pennucci}, T. and {Ransom}, S.~M. and
         {Roberts}, M.~S.~E. and {Hessels}, J.~W.~T.},
        title = "{A two-solar-mass neutron star measured using Shapiro delay}",
      journal = {\nat},
         year = 2010,
        month = oct,
       volume = {467},
       number = {7319},
        pages = {1081-1083},
          doi = {10.1038/nature09466},
archivePrefix = {arXiv},
       eprint = {1010.5788},
 primaryClass = {astro-ph.HE},
       adsurl = {https://ui.adsabs.harvard.edu/abs/2010Natur.467.1081D}
}

@Article{particles3040043,
AUTHOR = {Thapa, Vivek Baruah and Sinha, Monika and Li, Jia Jie and Sedrakian, Armen},
TITLE = "{Equation of state of strongly magnetized matter with hyperons and $\Delta$-resonances}",
JOURNAL = {Particles},
VOLUME = {3},
YEAR = {2020},
NUMBER = {4},
PAGES = {660--675},
URL = {https://www.mdpi.com/2571-712X/3/4/43},
ISSN = {2571-712X},
archivePrefix = {arXiv},
       eprint = {2010.00981},
 primaryClass = {hep-ph},
       adsurl = {https://ui.adsabs.harvard.edu/abs/2020arXiv201000981B},
DOI = {10.3390/particles3040043}
}

@ARTICLE{Sedrakian2007PrPNP,
       author = {{Sedrakian}, Armen},
        title = "{The physics of dense hadronic matter and compact stars}",
      journal = {Progress in Particle and Nuclear Physics},
         year = 2007,
        month = jan,
       volume = {58},
       number = {1},
        pages = {168-246},
          doi = {10.1016/j.ppnp.2006.02.002},
archivePrefix = {arXiv},
       eprint = {nucl-th/0601086},
 primaryClass = {nucl-th},
       adsurl = {https://ui.adsabs.harvard.edu/abs/2007PrPNP..58..168S}
}

@ARTICLE{1997NuPhA.625..435B,
       author = {{Balberg}, Shmuel and {Gal}, Avraham},
        title = "{An effective equation of state for dense matter with strangeness}",
      journal = {\nphysa},
         year = 1997,
        month = feb,
       volume = {625},
        pages = {435-472},
          doi = {10.1016/S0375-9474(97)81465-0},
archivePrefix = {arXiv},
       eprint = {nucl-th/9704013},
 primaryClass = {nucl-th},
       adsurl = {https://ui.adsabs.harvard.edu/abs/1997NuPhA.625..435B}
}

@article{Arzoumanian_2018,
	doi = {10.3847/1538-4365/aab5b0},
	url = {https://doi.org/10.3847/1538-4365/aab5b0},
	year = 2018,
	month = {apr},
	publisher = {American Astronomical Society},
	volume = {235},
	number = {2},
	pages = {37},
	author = {Zaven Arzoumanian and Adam Brazier and Sarah Burke-Spolaor and et al.},
	title = {The {NANOGrav} 11-year Data Set: High-precision Timing of 45 Millisecond Pulsars},
	journal = {The Astrophysical Journal Supplement Series}
}

@article{Aad_2023_1,
   title={Combination of searches for invisible decays of the Higgs boson using 139 fb$^{-1}$ of proton-proton collision data at $\sqrt{s}=13$ TeV collected with the ATLAS experiment},
   volume={842},
   ISSN={0370-2693},
   url={http://dx.doi.org/10.1016/j.physletb.2023.137963},
   DOI={10.1016/j.physletb.2023.137963},
   journal={Physics Letters B},
   publisher={Elsevier BV},
   author={{ATLAS Collaboration}},
   year={2023},
   month=jul,
   pages={137963}
}

@article{Aad_2023_2,
   title={Evidence of off-shell Higgs boson production from ZZ leptonic decay channels and constraints on its total width with the ATLAS detector},
   volume={846},
   ISSN={0370-2693},
   url={http://dx.doi.org/10.1016/j.physletb.2023.138223},
   DOI={10.1016/j.physletb.2023.138223},
   journal={Physics Letters B},
   publisher={Elsevier BV},
   author={{ATLAS Collaboration}},
   year={2023},
   month=nov,
   pages={138223}
}

@article{Parmar_2025,
   title={Exploring the
                    <mml:math xmlns:mml=“http://www.w3.org/1998/Math/MathML” display=“inline”>
                      <mml:mi mathvariant=“normal”>Δ</mml:mi>
                    </mml:math>
                    -resonance in neutron stars: Implications from astrophysical and nuclear observations},
   volume={112},
   ISSN={2470-0029},
   url={http://dx.doi.org/10.1103/643w-c2ly},
   DOI={10.1103/643w-c2ly},
   number={2},
   journal={Physical Review D},
   publisher={American Physical Society (APS)},
   author={Parmar, Vishal and Thapa, Vivek Baruah and Sinha, Monika and Bombaci, Ignazio},
   year={2025},
   month=July }

@article{Thapa_2021,
   title={Massive 
<mml:math xmlns:mml=“http://www.w3.org/1998/Math/MathML” display=“inline”><mml:mi mathvariant=“normal”>Δ</mml:mi></mml:math>
-resonance admixed hypernuclear stars with antikaon condensations},
   volume={103},
   ISSN={2470-0029},
   url={http://dx.doi.org/10.1103/PhysRevD.103.063004},
   DOI={10.1103/physrevd.103.063004},
   number={6},
   journal={Physical Review D},
   publisher={American Physical Society (APS)},
   author={Thapa, Vivek Baruah and Sinha, Monika and Li, Jia Jie and Sedrakian, Armen},
   year={2021},
   month=Mar }

@article{Malik_2022,
   title={Bayesian inference of signatures of hyperons inside neutron stars},
   volume={106},
   ISSN={2470-0029},
   url={http://dx.doi.org/10.1103/PhysRevD.106.063024},
   DOI={10.1103/physrevd.106.063024},
   number={6},
   journal={Physical Review D},
   publisher={American Physical Society (APS)},
   author={Malik, Tuhin and Providência, Constança},
   year={2022},
   month=Sept }

@article{Malikk_2022,
   title={Relativistic Description of Dense Matter Equation of State and Compatibility with Neutron Star Observables: A Bayesian Approach},
   volume={930},
   ISSN={1538-4357},
   url={http://dx.doi.org/10.3847/1538-4357/ac5d3c},
   DOI={10.3847/1538-4357/ac5d3c},
   number={1},
   journal={The Astrophysical Journal},
   publisher={American Astronomical Society},
   author={Malik, Tuhin and Ferreira, Márcio and Agrawal, B. K. and Providência, Constança},
   year={2022},
   month=Apr, pages={17} }

@article{Beznogov_2023,
   title={Bayesian inference of the dense matter equation of state built upon covariant density functionals},
   volume={107},
   ISSN={2469-9993},
   url={http://dx.doi.org/10.1103/PhysRevC.107.045803},
   DOI={10.1103/physrevc.107.045803},
   number={4},
   journal={Physical Review C},
   publisher={American Physical Society (APS)},
   author={Beznogov, Mikhail V. and Raduta, Adriana R.},
   year={2023},
   month=Apr}

@ARTICLE{Bonanno2012A&A,
       author = {{Bonanno}, Luca and {Sedrakian}, Armen},
        title = "{Composition and stability of hybrid stars with hyperons and quark color-superconductivity}",
      journal = {\aap},
         year = 2012,
        month = mar,
       volume = {539},
          eid = {A16},
        pages = {A16},
          doi = {10.1051/0004-6361/201117832},
archivePrefix = {arXiv},
       eprint = {1108.0559},
 primaryClass = {astro-ph.SR},
       adsurl = {https://ui.adsabs.harvard.edu/abs/2012A&A...539A..16B}
}

@article{Colucci_PRC_2013,
  title = {Equation of state of hypernuclear matter: Impact of hyperon--scalar-meson couplings},
  author = {Colucci, Giuseppe and Sedrakian, Armen},
  journal = {Phys. Rev. C},
  volume = {87},
  issue = {5},
  pages = {055806},
  numpages = {10},
  year = {2013},
  month = {May},
  publisher = {American Physical Society},
  doi = {10.1103/PhysRevC.87.055806},
  url = {https://link.aps.org/doi/10.1103/PhysRevC.87.055806}
}

@article{Oertel_RMP_2017,
      author         = "Oertel, M. and Hempel, M. and Klahn, T. and Typel, S.",
      title          = "{Equations of state for supernovae and compact stars}",
      journal        = "Rev. Mod. Phys.",
      volume         = "89",
      year           = "2017",
      number         = "1",
      pages          = "015007",
      doi            = "10.1103/RevModPhys.89.015007",
      SLACcitation   = "%%CITATION = ARXIV:1610.03361;%%"
}

@ARTICLE{Antoniadis_2013,
   author = {{Antoniadis}, J. and {Freire}, P.~C.~C. and {Wex}, N. and {Tauris}, T.~M. and 
	{Lynch}, R.~S. and {van Kerkwijk}, M.~H. and {Kramer}, M. and 
	{Bassa}, C. and {Dhillon}, V.~S. and {Driebe}, T. and {Hessels}, J.~W.~T. and 
	{Kaspi}, V.~M. and {Kondratiev}, V.~I. and {Langer}, N. and 
	{Marsh}, T.~R. and {McLaughlin}, M.~A. and {Pennucci}, T.~T. and 
	{Ransom}, S.~M. and {Stairs}, I.~H. and {van Leeuwen}, J. and 
	{Verbiest}, J.~P.~W. and {Whelan}, D.~G.},
    title = "{A Massive Pulsar in a Compact Relativistic Binary}",
  journal = {Science},
archivePrefix = "arXiv",
   eprint = {1304.6875},
 primaryClass = "astro-ph.HE",
     year = 2013,
   no-month = apr,
   volume = 340,
    pages = {448},
      doi = {10.1126/science.1233232},
   adsurl = {http://adsabs.harvard.edu/abs/2013Sci...340..448A}
}

@article{Drago_PRC_2014,
  title = {Early appearance of $\ensuremath{\Delta}$ isobars in neutron stars},
  author = {Drago, Alessandro and Lavagno, Andrea and Pagliara, Giuseppe and Pigato, Daniele},
  journal = {Phys. Rev. C},
  volume = {90},
  issue = {6},
  pages = {065809},
  numpages = {6},
  year = {2014},
  month = {Dec},
  publisher = {American Physical Society},
  doi = {10.1103/PhysRevC.90.065809},
  url = {https://link.aps.org/doi/10.1103/PhysRevC.90.065809}
}

@article{Cai_PRC_2015,
  title = {Critical density and impact of $\mathrm{\ensuremath{\Delta}}(1232)$ resonance formation in neutron stars},
  author = {Cai, Bao-Jun and Fattoyev, Farrukh J. and Li, Bao-An and Newton, William G.},
  journal = {Phys. Rev. C},
  volume = {92},
  issue = {1},
  pages = {015802},
  numpages = {11},
  year = {2015},
  month = {Jul},
  publisher = {American Physical Society},
  doi = {10.1103/PhysRevC.92.015802},
  url = {https://link.aps.org/doi/10.1103/PhysRevC.92.015802}
}

@article{Zhu_PRC_2016,
  title = {$\mathrm{\ensuremath{\Delta}}(1232)$ effects in density-dependent relativistic Hartree-Fock theory and neutron stars},
  author = {Zhu, Zhen-Yu and Li, Ang and Hu, Jin-Niu and Sagawa, Hiroyuki},
  journal = {Phys. Rev. C},
  volume = {94},
  issue = {4},
  pages = {045803},
  numpages = {13},
  year = {2016},
  month = {Oct},
  publisher = {American Physical Society},
  doi = {10.1103/PhysRevC.94.045803},
  url = {https://link.aps.org/doi/10.1103/PhysRevC.94.045803}
}

@article{Weissenborn2012a,
  title={Hyperons and massive neutron stars: The role of hyperon potentials},
  author={Weissenborn, Simon and Chatterjee, Debarati and Schaffner-Bielich, Juergen},
  journal={Nucl. Phys. A},
  volume={881},
  pages={62--77},
  year={2012},
  publisher={Elsevier}
}

@article{Chatterjee2015,
      author         = "Chatterjee, Debarati and Vidana, Isaac",
      title          = "{Do hyperons exist in the interior of neutron stars?}",
      journal        = "Eur. Phys. J.",
      volume         = "A52",
      year           = "2016",
      number         = "2",
      pages          = "29",
      doi            = "10.1140/epja/i2016-16029-x",
      eprint         = "1510.06306",
      archivePrefix  = "arXiv",
      primaryClass   = "nucl-th",
      SLACcitation   = "%%CITATION = ARXIV:1510.06306;%%"
}

@article{Oertel2015,
  title={Hyperons in neutron star matter within relativistic mean-field models},
  author={Oertel, M and Provid{\^e}ncia, C and Gulminelli, F and Raduta, Ad R},
  journal={J. Phys. G},
  volume={42},
  number={7},
  pages={075202},
  year={2015},
  publisher={IOP Publishing}
}

@article{PhysRevX.9.011001,
  title = {Properties of the Binary Neutron Star Merger GW170817},
  author = {Abbott, B. P. and Abbott, R. and Abbott, T. D. and Acernese, F. and Ackley, K. and Adams, C. and others},
  collaboration = {LIGO Scientific Collaboration and Virgo Collaboration},
  journal = {Phys. Rev. X},
  volume = {9},
  issue = {1},
  pages = {011001},
  numpages = {32},
  year = {2019},
  month = {Jan},
  publisher = {American Physical Society},
  doi = {10.1103/PhysRevX.9.011001},
  url = {https://link.aps.org/doi/10.1103/PhysRevX.9.011001}
}

@article{Shahrbaf_2026_probing,
   title={Probing strange dark matter through f-mode oscillations of neutron stars with hyperons and quark matter},
   volume={2026},
   ISSN={1475-7516},
   url={http://dx.doi.org/10.1088/1475-7516/2026/03/017},
   DOI={10.1088/1475-7516/2026/03/017},
   number={03},
   journal={Journal of Cosmology and Astroparticle Physics},
   publisher={IOP Publishing},
   author={Shahrbaf, Mahboubeh and Thakur, Prashant and Rafiei Karkevandi, Davood},
   year={2026},
   month=Mar, pages={017} }

@article{Shahrbaf_2026_observational,
   title={Observational probes of the neutron star equation of state with hyperons, bosonic dark matter, and quark matter},
   volume={706},
   ISSN={1432-0746},
   url={http://dx.doi.org/10.1051/0004-6361/202556315},
   DOI={10.1051/0004-6361/202556315},
   journal={\aap},
   publisher={EDP Sciences},
   author={Shahrbaf, Mahboubeh and Rafiei Karkevandi, Davood and Ayriyan, Alexander and Typel, Stefan},
   year={2026},
   month=Feb, pages={A203} }

@article{McKeen_2021_Mirror,
  title = {Neutron Star Internal Heating Constraints on Mirror Matter},
  author = {McKeen, David and Pospelov, Maxim and Raj, Nirmal},
  journal = {Phys. Rev. Lett.},
  volume = {127},
  issue = {6},
  pages = {061805},
  numpages = {6},
  year = {2021},
  month = {Aug},
  publisher = {American Physical Society},
  doi = {10.1103/PhysRevLett.127.061805},
  url = {https://link.aps.org/doi/10.1103/PhysRevLett.127.061805}
}

@ARTICLE{2022ApJ...939L..34A,
       author = {{Altiparmak}, Sinan and {Ecker}, Christian and {Rezzolla}, Luciano},
        title = "{On the Sound Speed in Neutron Stars}",
      journal = {\apjl},
         year = 2022,
        month = nov,
       volume = {939},
       number = {2},
          eid = {L34},
        pages = {L34},
          doi = {10.3847/2041-8213/ac9b2a},
archivePrefix = {arXiv},
       eprint = {2203.14974},
 primaryClass = {astro-ph.HE},
       adsurl = {https://ui.adsabs.harvard.edu/abs/2022ApJ...939L..34A}
}

\end{document}